\pdfoutput=1

\documentclass[11pt]{article}

\usepackage{acl}

\usepackage{times}
\usepackage{latexsym}

\usepackage{amsmath,amsfonts,bm}

\def\eqref#1{equation~\ref{#1}}

\def\1{\bm{1}}

\def\vh{{\bm{h}}}

\def\mH{{\bm{H}}}

\def\mK{{\bm{K}}}

\def\mQ{{\bm{Q}}}

\def\mV{{\bm{V}}}
\def\mW{{\bm{W}}}

\DeclareMathAlphabet{\mathsfit}{\encodingdefault}{\sfdefault}{m}{sl}
\SetMathAlphabet{\mathsfit}{bold}{\encodingdefault}{\sfdefault}{bx}{n}

\usepackage[T1]{fontenc}

\usepackage[utf8]{inputenc}

\usepackage{microtype}

\usepackage{inconsolata}

\usepackage{graphicx}
\usepackage{caption}
\usepackage{subcaption}

\usepackage{authblk}

\title{A Critical Study of What Code-LLMs (Do Not) Learn}

\author[1,a]{Abhinav Anand}
\author[1,a]{Shweta Verma}
\author[2,b]{Krishna Narasimhan\thanks{Work conducted while afiliated with Technische Universität Darmstadt.}}
\author[1,3,4,c]{Mira Mezini}
\affil[1]{Technische Universität Darmstadt}
\affil[2]{AI Quality \& Testing Hub}
\affil[3]{Hessian Center for Artificial Intelligence (hessian.AI)}
\affil[4]{National Research Center for Applied Cybersecurity ATHENE}
\affil[a]{\texttt {\{abhinav.anand, shweta.verma\}@tu-darmstadt.de}}
\affil[b]{\texttt{k.narasimhan@aiqualityhub.com}}
\affil[c]{\texttt{mezini@cs.tu-darmstadt.de}}

\begin{document}
\maketitle
\begin{abstract}

Large Language Models trained on code corpora (code-LLMs) have demonstrated impressive performance in various coding assistance tasks. However, despite their increased size and training dataset, code-LLMs still have limitations such as suggesting codes with syntactic errors, variable misuse etc. Some studies argue that code-LLMs perform well on coding tasks because they use self-attention and hidden representations to encode relations among input tokens. However, previous works have not studied what code properties are not encoded by code-LLMs. In this paper, we conduct a fine-grained analysis of attention maps and hidden representations of code-LLMs. Our study indicates that code-LLMs only encode relations among specific subsets of input tokens. Specifically, by categorizing input tokens into syntactic tokens and identifiers, we found that models encode relations among syntactic tokens and among identifiers, but they fail to encode relations between syntactic tokens and identifiers. We also found that fine-tuned models encode these relations poorly compared to their pre-trained counterparts. Additionally, larger models with billions of parameters encode significantly less information about code than models with only a few hundred million parameters.

\end{abstract}

\section{Introduction}
Code-LLMs (cLLMs) are Transformer models \citep{transformer} trained on a large corpus of code and natural language - programming language (NL-PL) pairs. These models are used, either in a zero-shot manner or after fine-tuning, for coding assistance tasks, including code summarization, code retrieval, code completion, code generation, and program repair \citep{survey_xu}. 

While the performance of models on benchmarks has significantly improved in the past few years, there are still issues with performance in real-world settings.
Code generated by
models has compilation errors due to syntactical mistakes \citep{coderl}, semantic errors like random identifiers \citep{graphcodebert}, and can invoke undefined or out-of-scope 
functions, variables and attributes \citep{codex}. Some studies suggest that models do not generalize well \citep{simscood, hellendoorn}, learn shortcuts \citep{code_summay, simple_rabin}, 
and memorize training inputs \citep{mem_rabin, mem_yang}. 
To understand the cause of these issues, it is imperative to understand which code properties are used by cLLMs for prediction and generation and which are not encoded by cLLM.
But the black-box nature of neural networks 
makes this understanding a challenging task.

Prior studies have 
used
attention analysis \citep{icse_capture} and probing on hidden representation \citep{belinkov} to 
study what cLLMs encode. Some of these studies show that models can learn the syntactic and semantic structure of code  \citep{icse_capture, troshin, astprobe} and understand code logic \citep{logic_probe}. 
However, they rely on non-systematically validated assumptions. For example, studies on attention analysis set an arbitrary attention threshold of 0.3. The studies which probe hidden representation of code 
models assume a linear encoding of information. The effect of these assumptions has hitherto remained unstudied. Further, these studies do not evaluate which code properties are not encoded by cLLMs. 
In this paper, we make two important contributions to advance the state of the art in the interpretability of cLLMs.

First, we perform a systematic analysis of assumptions in previous work and 
show that they can lead to misleading conclusions. Specifically, we examine the influence of the attention 
threshold and evaluation metric on attention analysis, and for probing on hidden representation, we explore whether the code relations among tokens are encoded linearly or non-linearly. 
To avoid several limitations of 
classifier and structural probing methods \citep{tale, control_task, belinkov}, we perform probing of hidden representation without any additional classifier layers or parameters. Based on our observations, we make some new suggestions for experimental setup of analysis of attention maps and hidden representation.

Second, armed with our insights from the first analysis, we set up and perform a fine-grained analysis of attention and hidden representation of cLLMs at the code token level to critically examine what they learn and do not learn. Previous studies examining 
the code comprehension ability of cLLMs have analyzed all input tokens together, without distinguishing between different categories of code tokens such as identifiers 
(e.g., function names, variables) and syntactic tokens (e.g., keywords, operators, parentheses).  To investigate whether there are specific relations that cLLMs fail to encode, we separately analyze 
the syntactic-syntactic, identifier-identifier, and syntactic-identifier relations that are encoded in the self-attention values and hidden representations. 

There are 
different types of relations between code tokens, including relations in an abstract syntax tree (AST), as well as, data flow or control flow relations between code blocks. 
Similar to \citet{icse_capture}, we focus primarily on syntactic relations in the AST and create a syntax graph with edges between code tokens within a motif structure (Figure \ref{fig: ast}). 
But such a syntax graph does not encompass all the relations among identifiers, in particular how values flow from one variable to another. 
Thus, we extend the study to data-flow relations and create a data flow graph (DFG) with edges among related variables following \citet{graphcodebert}. 

We perform attention analysis to study whether a token pays attention to related tokens and analysis of hidden representation to study the information encoded by the model in the vector representation of a token. To study information encoded in hidden representations, we take hidden representations of pairs of tokens and evaluate if the information encoded by the model is sufficient to predict the relation between these two tokens. Specifically, we evaluate with respect to predicting edges in a DFG and sibling and distance prediction in an AST.

\begin{figure}[h]
\begin{center}
    \begin{subfigure}[b]{0.3\textwidth}
    \begin{center}
    \includegraphics[width=\linewidth]{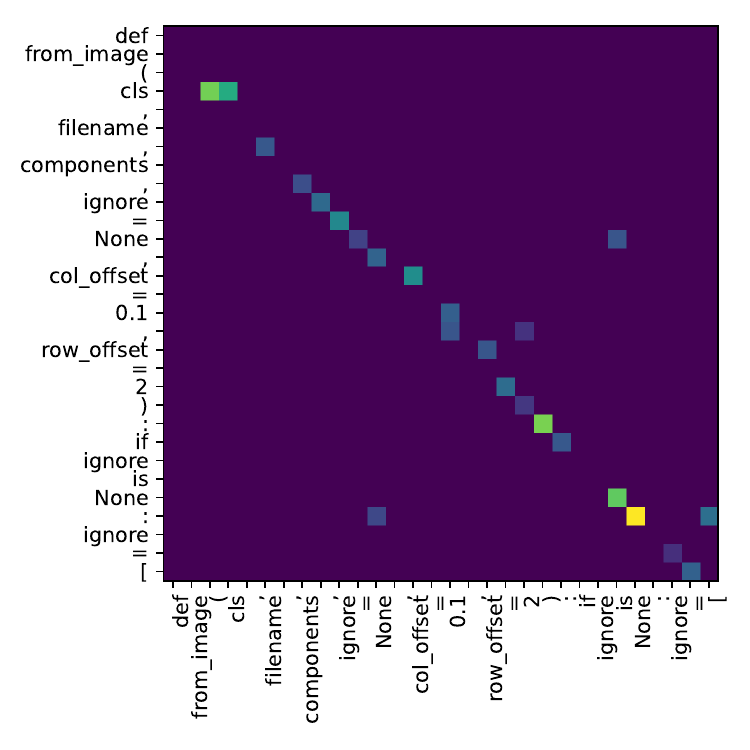}
    \end{center}
    \label{fig: att_best_p}
    \end{subfigure}
    \hfill
    \begin{subfigure}[b]{0.3\textwidth}
    \begin{center}
    \includegraphics[width=\linewidth]{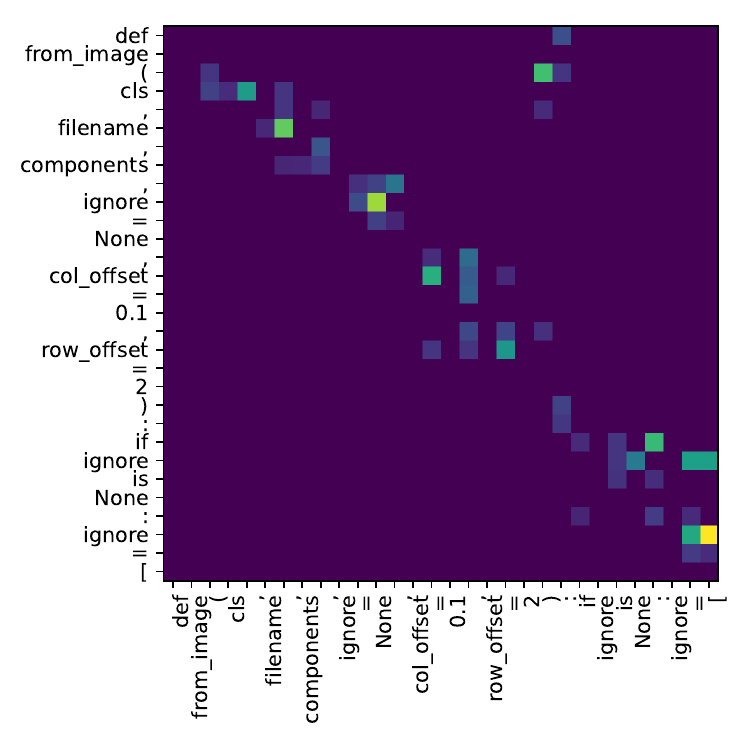}
    \end{center}
    \label{fig: att_best_f}
    \end{subfigure}
\end{center}
\caption{Attention map for head with best precision (head 1) \textbf{(top)} and head with best f-score (head 2) \textbf{(bottom)} of layer 9 of CodeBERT for first 30 tokens of a python code (see Figure \ref{fig: code} for code). The head with best precision mostly encodes next-token attention, while head with best f-score encodes more complex relation.}
\label{fig: explanation}
\end{figure}

We study models with 110M to 3.7B parameters with different architectures, pre-training objectives, and training datasets \footnote{The code is available at \url{https://github.com/stg-tud/code-LLM-critical-evaluation}.}. In summary,

\begin{itemize}
    \item We provide evidence that prior work often made incorrect assumptions in their experimental settings, which led to misleading conclusions. In particular, previous works on attention analysis assume an attention threshold of 0.3 and study heads with best precision (shown in Figure \ref{fig: explanation}). Also, the studies on hidden representation assume linear encoding of information in hidden representation.
    \item The attention maps of cLLMs fall short in encoding syntactic-identifier relations, while they do encode syntactic-syntactic and identifier-identifier relations. Also, the hidden representations of cLLMs do not encode sufficient information to discriminate between different identifier types and to understand subtle syntactical differences.
    \item We show that the issues of cLLMs with encoding code syntax persists for big models with significantly increased number of parameters or for models that are fine-tuned on specific tasks. In fact, we observe a reduction in encoding code syntax and even data-flow relations with large size and fine-tuning.
\end{itemize}


\section{Related Work}
Several studies have provided some possible explaination of the working of cLLMs. \citet{conterfactual} and \citet{distractors} used input perturbation, while, \citet{shap_liu} used backpropagation to find the most relevant input tokens. \citet{att_aggr} created an aggregated attention graph and studied its application to the \texttt{VarMiuse} task. \citet{icse_capture} performed attention analysis and probing with structural probes \citep{sp}. \citet{astprobe} used structural probe to create binarized AST from hidden representations. 

Probing classifiers have been used to test syntax and semantic understanding \citep{probe_karmakar, troshin, probe_ahmed}, the effect of positional embeddings \citep{probe_yang}, relation between self-attention and distance in AST \citep{cat_probe} and logic understanding \citep{logic_probe}.

Other studies established correlations between input tokens, model output, and self-attention. \citet{autofocus} created an attention-based discriminative score to rank input tokens and studied the impact of high-ranked tokens on output. Attention-based token selection was utilized by \citet{dietcode} to simplify the input program of CodeBERT \citep{codebert}. \citet{simple_rabin} and \citet{syntax_rabin} simplified the input program while preserving the output and showed that the percentage of common tokens between attention and reduced input programs is typically high.


\textbf{Our Work} studies the limitations of code models in encoding code structure which has hitherto remained unexplored. Our study spanning multiple transformer architectures, sizes and training objectives demonstrate a significant gap in encoding some code properties. This gap could be a possible explanation for poor performance of cLLMs on real-world tasks \citep{hellendoorn}. 


\section{Experiments}
In this section, we elaborate on the experiments that we performed to analyze self-attention and the hidden representation of cLLMs. For attention analysis, we compare the self-attention of models with the motif structure in a program's AST and DFG. For hidden representations, we perform probing without classifiers using DirectProbe \citep{directprobe}. We provide details on AST, DFG, DirectProbe, and motif structure in Appendix \ref{background}.

\subsection{Models and Dataset}
We analyze 
a wide range of pre-trained and fine-tuned models. The parameters range from 110M to 3.7B. The investigated models also have different architectures, training datasets, and objectives. 

Among the subjects there are the encoder-only models such as CodeBERT \citep{codebert} and GraphCodeBERT \citep{graphcodebert}), encoder-decoder models such as CodeT5 \citep{codet5}, PLBART \citep{plbart} and CodeT5+ \citep{codet5p}, and decoder-only models. CodeGen \citep{codegen} is a decoder-only model trained with fill-in-the-middle objective \citep{fim} for bi-directional context while UnixCoder with encode-decoder architecture \citep{unixcoder} has a UniLM-style \citep{unilm} training. 

We also investigate models with different objectives.
CodeT5-musu \citep{codet5} is fine-tuned for summarization tasks, CodeT5+220Mbi \citep{codet5p} can be used in a zero-shot manner for summarization and retrieval tasks, and CodeRL \citep{coderl} is a larger CodeT5 model (CodeT5\_lntp) trained for code generation in an actor-critic setup using test cases for reward.    

For our experiments, we randomly sampled 3000 Python codes from the test set of CodeSearchNet dataset \citep{csn} after removing docstrings and comments. More details about the models and data preparation are presented in Appendix \ref{model} and Appendix \ref{dataset} respectively.  

\subsection{Attention Analysis}
\subsubsection{Setup} \label{setup}
\textbf{Model graph.}
The attention map of a head is a $n*n$ matrix ($n$ is the number of input tokens). The elements of the matrix represent the significance each token attributes to other tokens. We consider the matrix as the adjacency matrix of a graph with input tokens corresponding to nodes and attention values inducing an edge. Similar to previous works on attention analysis \citep{icse_capture, att_aggr}, we merge the sub-tokens of input code tokens by averaging their attention values.

We considered the edges of the model graphs as predictions and that of code graphs (defined later) as the ground truth in the computation of precision and recall.

\begin{figure}[htp]
\begin{center}
    \begin{subfigure}[b]{0.22\textwidth}
    \begin{center}
     \includegraphics[width=1\linewidth]{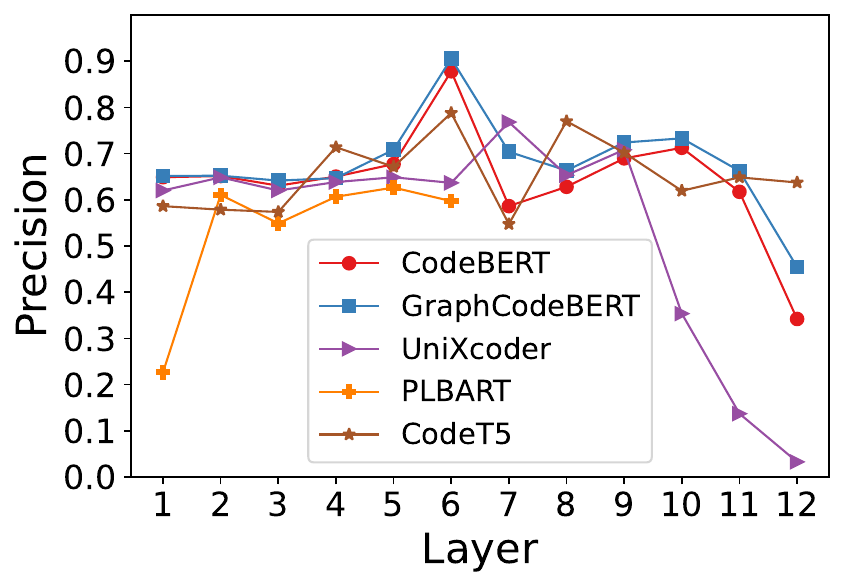}   
    \end{center}
    \end{subfigure}
    \hspace{1em}
    \begin{subfigure}[b]{0.22\textwidth}
    \begin{center}
    \includegraphics[width=1\linewidth]{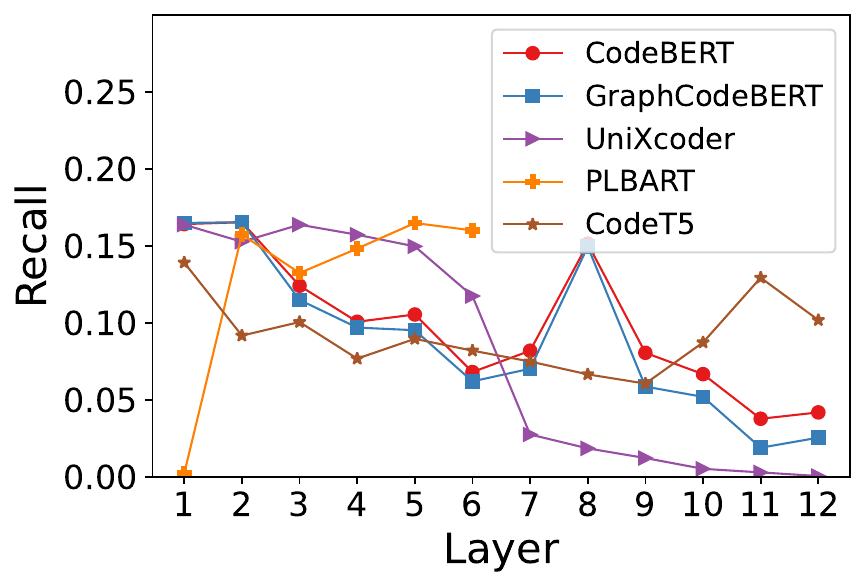}
    \end{center}
    \end{subfigure}
\end{center}
\caption{On comparing model graph with syntax graph with an attention threshold of 0.3, the precision (left) is high but the recall is very low (right).}
\label{fig: thr}
\end{figure}

Prior studies have typically set an arbitrary threshold of 0.3 for attention analysis and have excluded heads with very few attention values, usually less than 100, from the analysis \citep{icse_capture, bmeetsb}. This approach excludes more than 99.5\% of self-attention values (see Appendix \ref{distrib}), thereby skewing the conclusions drawn. For instance, \citet{icse_capture} reported high precision values, indicating that the majority of attention values correspond to relations in the AST. However, we observe a significantly reduced recall, as shown in Figure \ref{fig: thr}. The low recall shows that only a small proportion of syntactic relations are encoded in attention values greater than 0.3. Further, a code token is always syntactically related to the next token, unless there is a line break in between. Consequently, encoding next token attention results in high precision. As shown in Figure \ref{fig: explanation}, the heads with best precision often only encode next-token attention. On the other hand, heads with best f-score encode more relations such as attention paid to tokens other than the next-token. 

\begin{figure}[htp]
\centering
    \begin{subfigure}[b]{0.22\textwidth}
    \begin{center}
     \includegraphics[width=0.9\linewidth]{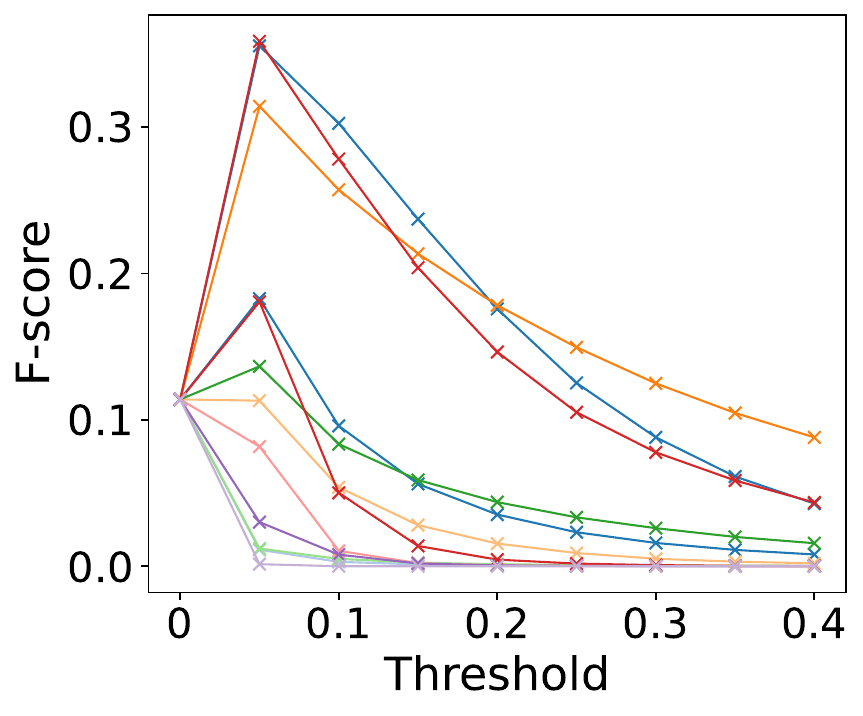}   
    \end{center}
    \caption{CodeBERT layer 6}
    \end{subfigure}
    \hfill
    \begin{subfigure}[b]{0.22\textwidth}
    \begin{center}
    \includegraphics[width=0.9\linewidth]{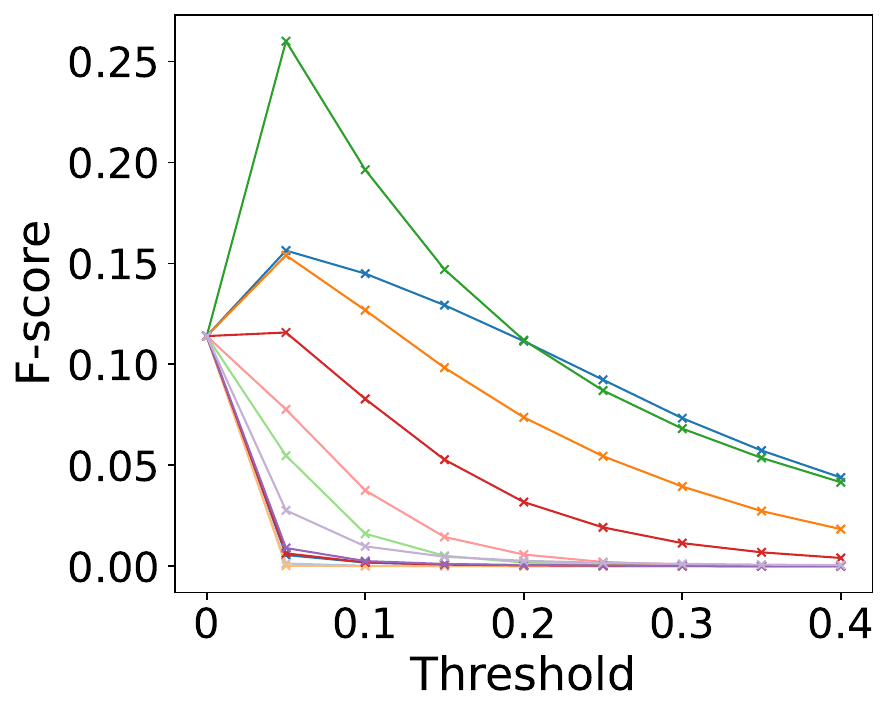}
    \end{center}
    \caption{CodeBERT layer 12}
    \end{subfigure}
    \hfill
    \begin{subfigure}[b]{0.22\textwidth}
    \begin{center}
    \includegraphics[width=0.9\linewidth]{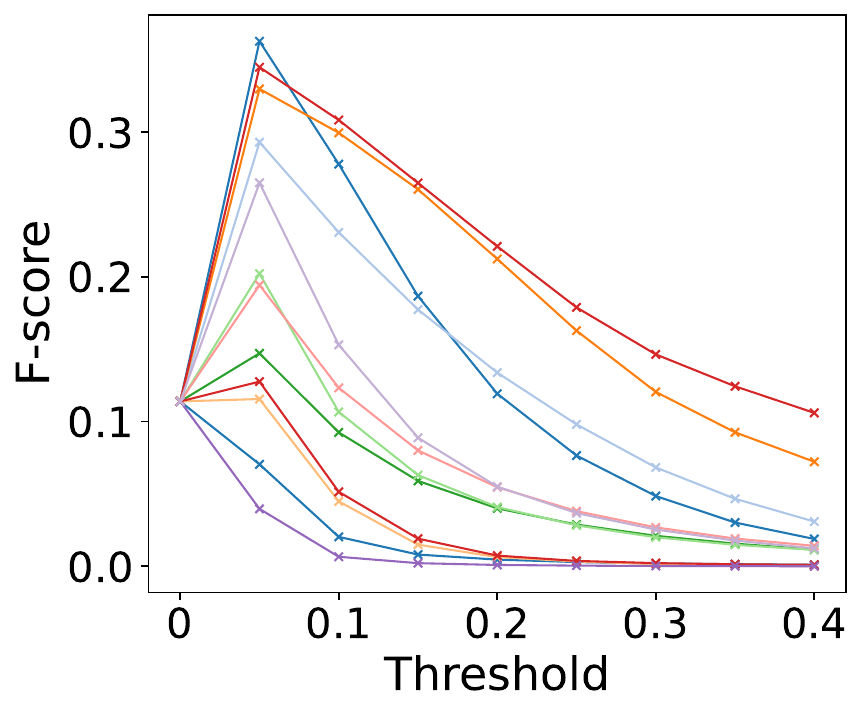}
    \end{center}
    \caption{CodeT5 layer 6}
    \end{subfigure}
        \hfill
    \begin{subfigure}[b]{0.22\textwidth}
    \begin{center}
    \includegraphics[width=0.9\linewidth]{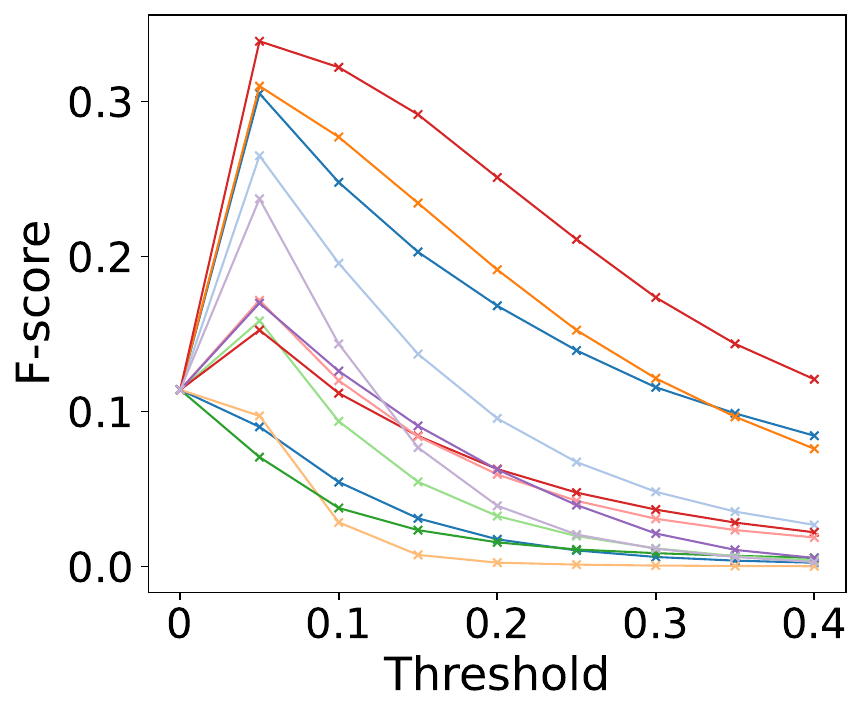}
    \end{center}
    \caption{CodeT5 layer 12}
    \end{subfigure}
\caption{ The plot illustrates F-score between model graph and syntax graph at different thresholds for all heads. Each curve in a plot represents one head. The plots for layer 6 and layer 12 of CodeBERT and CodeT5 are shown out of various models and layers evaluated at different thresholds. For most heads, F-score is highest at a threshold of 0.05 for all models.} 
\label{fig: fscr}
\end{figure}

So, to balance between precision and recall, we use F-score. We evaluate F-scores for all heads across various models and layers at different threshold values. As shown in Figure \ref{fig: fscr}, the highest F-score is achieved when using a threshold of 0.05. We use this threshold for all experiments.
Similar to previous works \citep{icse_capture}, we set all values below the threshold to 0 and those above to 1. That is, we don't weight the calculations with actual self-attention values. Such a weighting
will lower the precision and recall and increase the graph edit distance per node (Section \ref{analysis}). Setting values to 1 refers to the best-case scenario. Thus, the limitations documented in this work exist even in the best-case scenario. Weighing with original values will only make these limitations more stark without changing the conclusion.

\textbf{Code graphs.} We compare the \emph{model graph} with two \emph{code graphs}: the \emph{syntax graph}, representing relations in an AST, and the \emph{DFG graph}. 
The syntax graph comprises syntactic relations among all tokens, while the DFG comprises data flow relations among identifiers.
Following \citet{icse_capture}, we assume two tokens to have a syntactic relation if they exist in the same motif structure (see Appendix \ref{background}).  Since we want to study the encoding of syntactic-syntactic, identifier-identifier, and syntactic-identifier relations separately, we create a \emph{non-identifier graph} with the same nodes as the syntax graph but only encompassing AST relations between syntactic tokens.

\subsubsection{Analysis} \label{analysis}
For each model, we compare the model graph of a head with the code graphs in two ways. 

First, we compute the precision and recall between the set of edges in the model graph and the code graphs. We consider the edges of the code graphs as ground truth and those of the model graphs as predictions. For comparison across layers of a model, we select the heads with the highest F-score for each layer. 

Second, we calculate the graph edit distance (GED) \citep{ged} per node to quantify the similarity between code and model graphs. GED between two graphs $G_1$ and $G_2$ computes the cost of inserting, deleting, or substituting nodes and edges to transform $G_1$ into an isomorphic graph of $G_2$. Code graphs and model graphs share the same set of nodes and have only one edge type. So, we assign a cost of 1 for both edge deletion and insertion operations and 0 otherwise. In all calculations, we apply the operations to model graphs. We also calculate the GED between the model graph and the non-identifier graph. For GED calculations, we use the NetworkX package \citep{networkx}.

\subsection{Analysis of Hidden Representations} 
\subsubsection{Qualitative Analysis with t-SNE} \label{qualitative}
The hidden representation, $\vh_i^l$ of $i^{th}$ word at the output of layer $l$, is a $d$-dimensional vector. We use 
t-SNE \citep{tsne} -- a widely used technique to project high-dimensional data into a two-dimensional space while preserving the distance distribution between points - to qualitatively analyze the hidden representations in two settings.  

First, we study the distribution of hidden representations of different token types; to this end, 
we collect the hidden representations of code tokens of specific types from 100 programs, each having a minimum of 100 tokens. 

Second, we compare the distance distribution between tokens in an AST and between their hidden representations. In the AST, siblings have similar distance distribution. So, in t-SNE visualization of AST  tree distances, siblings cluster together. If the distance between hidden representations corresponds to the distance in the AST, hidden representations should also have a similar distance distribution. To this end, we construct distance matrices of both for randomly selected code samples.

\subsubsection{Probing on Hidden Representations}
\label{probing}
We use DirectProbe \citep{directprobe} to quantitatively evaluate the syntactic and data flow information encoded in hidden representations of each token for a given layer. We create datasets for each layer of the models we examined. Each data point is represented as $(\vh_i^l  * \vh_j^l): label_t$. $* \in \{concatenation, difference\}$ is an operation between hidden representations of tokens $i$ and $j$ of layer $l$. $t \in \{siblings, tree distance, data flow\}$ is a task to evaluate whether hidden representations encode information about the specific property. Each dataset is split in a $80:20$ ratio into training and test sets. The training set is used to create clusters for each label and the test set is used to evaluate the quality of clustering. 

Using $data flow$, we study whether data flow relations are encoded. Here, both $i$ and $j$ are identifiers, $label\in \{NoEdge, ComesFrom, ComputedFrom\}$ and $* = concatenation$. Using $siblings$ and $tree distance$, we study the encoding of relations in an AST. For both tasks, token $i$ is one of a subset of Python keywords (listed in Appendix \ref{directprobe}). In one set of experiments, (\texttt{Keyword-All}), token $j$ can be any other token. In another set, (\texttt{Keyword-Identifier}), token $j$ is an identifier. For the siblings task, $label\in \{sibling, not sibling\}$, where two tokens in the same motif structure are considered to be siblings, and $* = concatenation$. The minimum distance between two code tokens in an AST is $2$ while tokens far apart in an AST don't have any discriminative syntactic relations. So, for tree distance, we only consider $label\in \{2, 3, 4, 5, 6\}$. Moreover, \citet{bert_geom} showed that square of distance between two vectors, $(\vh_i^l-\vh_j^l)^T(\vh_i^l-\vh_j^l)$, corresponds to distance in a tree. Hence, we set $* = difference$ for the distance prediction task.

The tree distance between a keyword and an identifier denotes different identifier types and syntax structures. For instance, consider the statements of the form (a) \texttt{if var1:} and (b) \texttt{if var1 == var2:}. The tree distance between \texttt{if} and \texttt{var1} is 2 in (a) and 3 in (b). In a function declaration, the identifier types function name, parameters, and default parameters are, respectively, at a distance of $2$, $3$ and $4$ from \texttt{def}. Hence, if the hidden representations encode information about different identifier types and syntax, it follows that hidden representations of (\texttt{Keyword-Identifier}) pairs at a certain distance in AST must form separable clusters.   

\section{Results}
\begin{figure*}[htbp]
    \centering
    \begin{subfigure}[b]{0.99\textwidth}
     \includegraphics[height=0.165\linewidth]{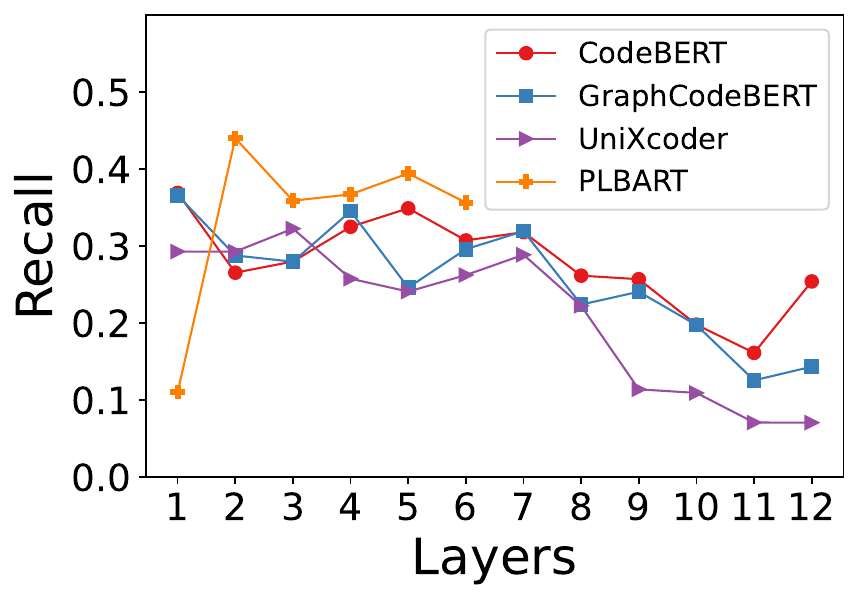} 
     \includegraphics[height=0.165\linewidth]{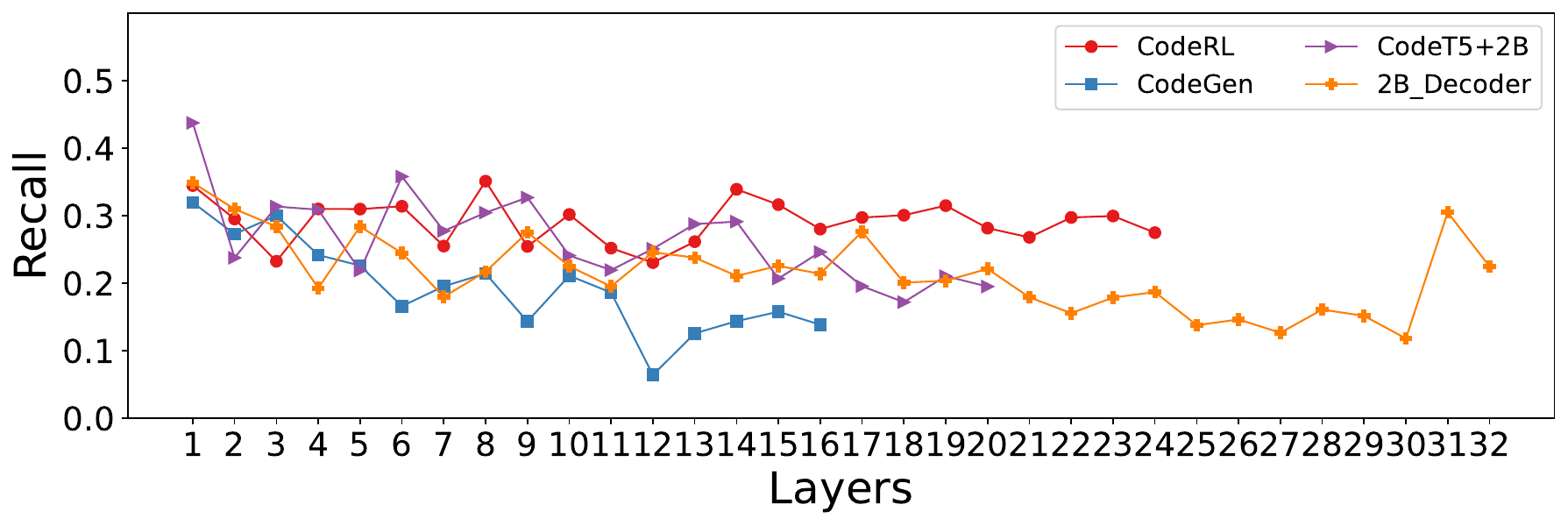} 
     \includegraphics[height=0.165\linewidth]{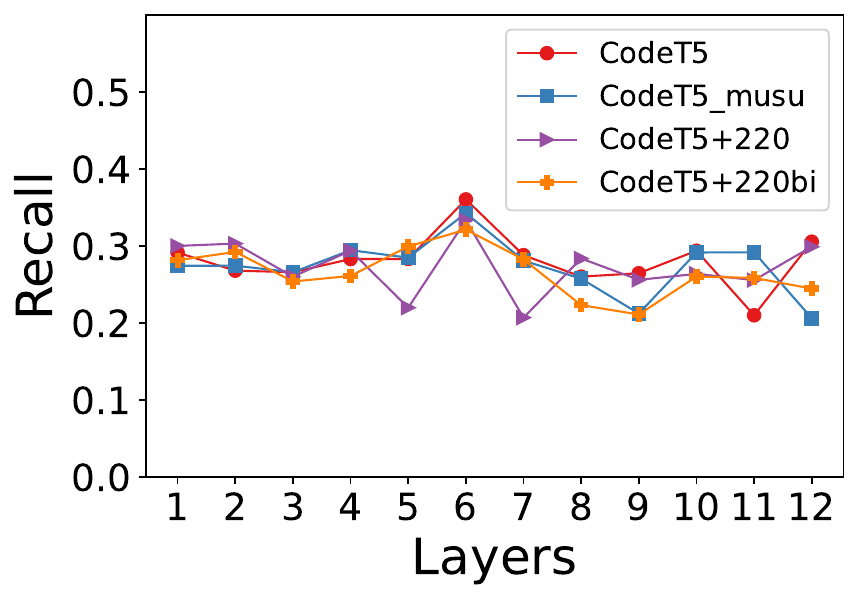} 
    \end{subfigure}
        \begin{subfigure}[b]{0.99\textwidth}
     \includegraphics[height=0.166\linewidth]{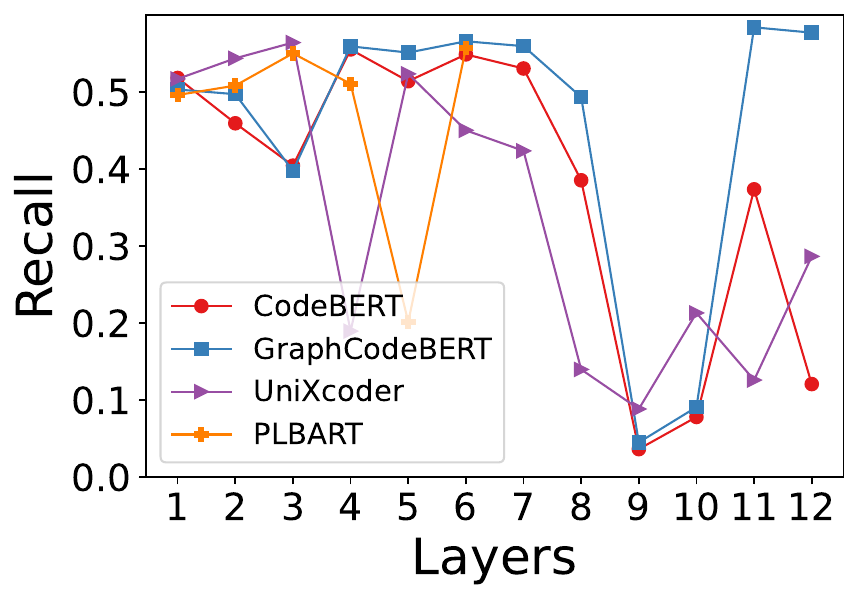} 
     \includegraphics[height=0.166\linewidth]{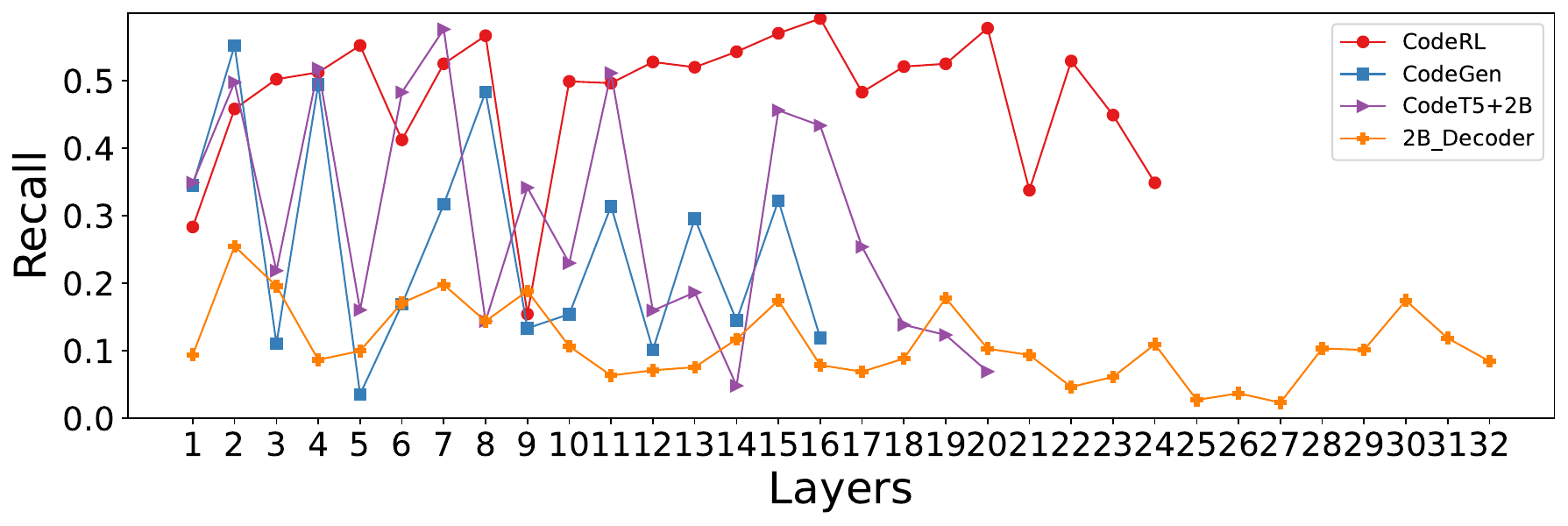} 
     \includegraphics[height=0.166\linewidth]{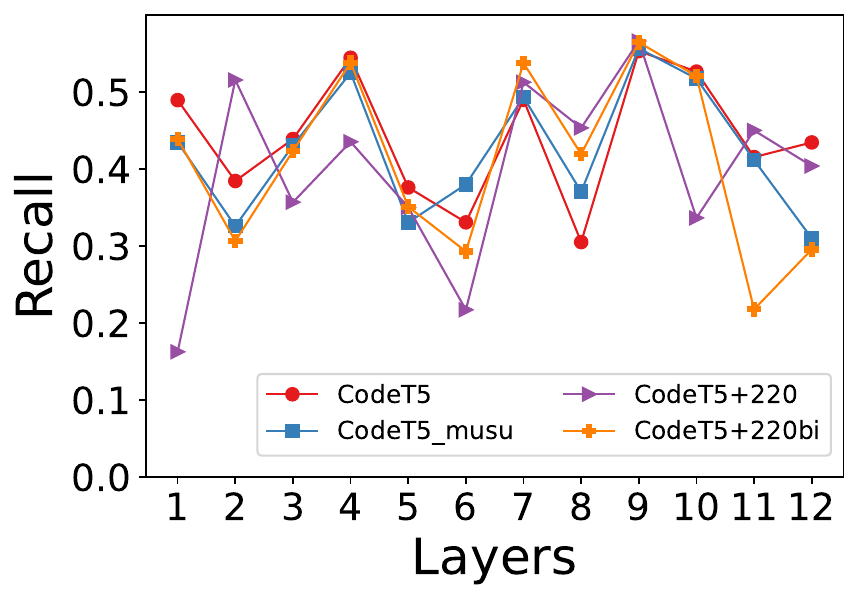} 
    \end{subfigure}
    \caption{Recall of model graphs with syntax graphs (top) and data flow graphs (bottom). The plots show irrespective of training-objectives, fine-tuning or larger sizes, the models do not encode more than 40\% of syntactic relations and around 55\% of data flow relations. Enc-Dec models encode syntactic relations much better in deeper layers.}
    \label{fig: recall}
\end{figure*}

\begin{figure*}[htbp]
\centering
    \begin{subfigure}[b]{0.99\textwidth}
     \includegraphics[height=0.169\linewidth]{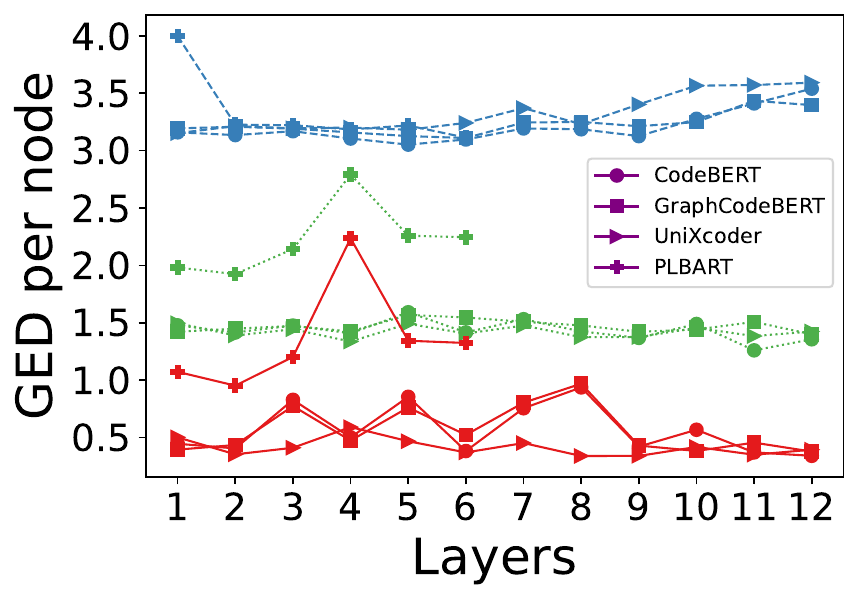} 
     \includegraphics[height=0.169\linewidth]{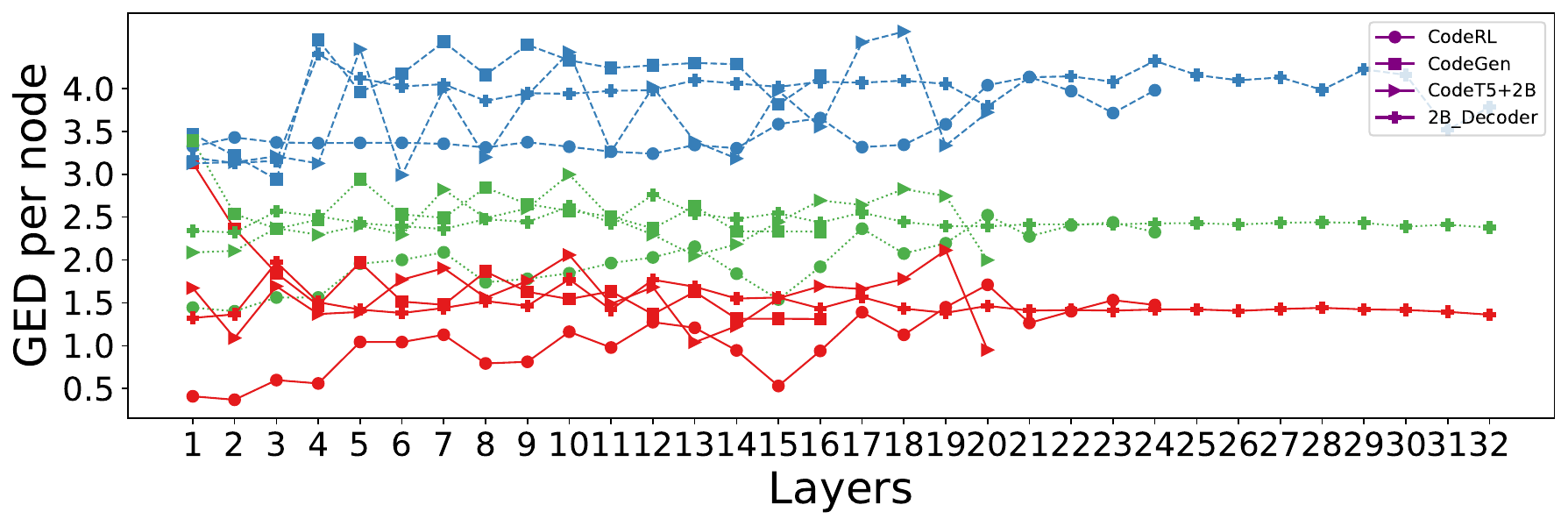}
     \includegraphics[height=0.169\linewidth]{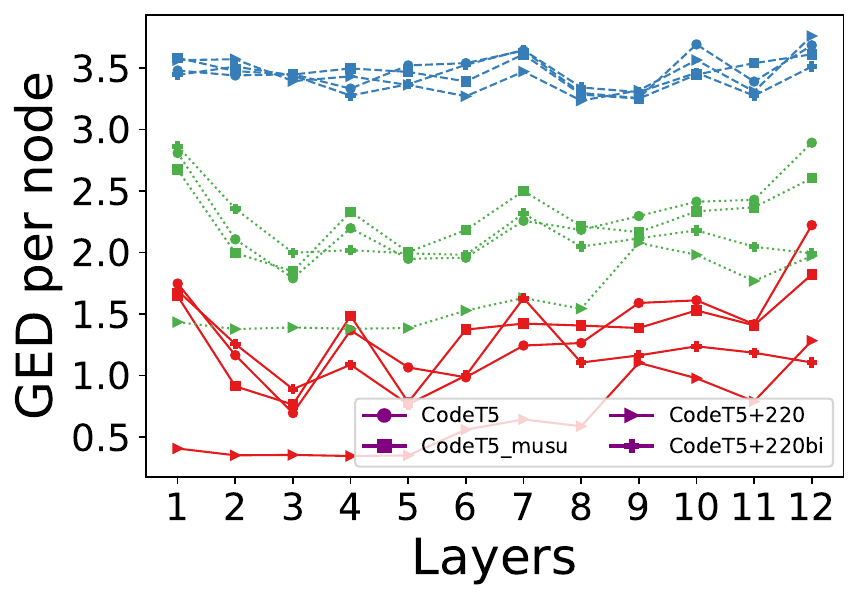} 
    \end{subfigure}
    \begin{subfigure}[b]{0.99\textwidth}
    \centering
    \includegraphics[width=0.478\textwidth]{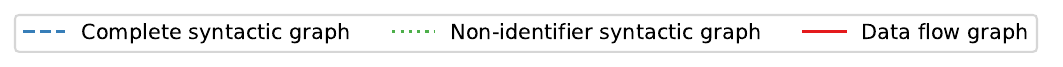} 
    \end{subfigure}
    \caption{Graph edit distance (GED) per node (lower value show higher similarity) of model graph from DFG, non-identifier syntax graph and complete syntax graph for various models. The gap between non-identifier and complete syntax graph shows that on introducing syntax-identifier edges the similarity reduces drastically and thus, these edges are not present in the model graphs. For very large models (center), even DFG edges are encoded poorly.}
    \label{fig: similarity}
\end{figure*}

\subsection{Attention Analysis}
In Figure \ref{fig: recall} we present the recall between model graphs and code graphs. We observe that different models encode code relations to varying degrees. Surprisingly, fine-tuned and larger models do not encode a higher proportion of code relations compared to smaller pre-trained models, even if they perform  
better on benchmarks.  
Similarly, the actor-critic training of CodeRL does not improve encoding of code relation compared to CodeT5\_lntp, even if it performs significantly better on code generation\cite{coderl}. Further, the decoder-only CodeGen model with 3.7B parameters barely encodes code-relations in deeper layers.  

We also find that the proportion of encoded relations degenerate in deeper layers of encoder-only models but not in encoder-decoder models. The degeneration in deeper layers of encoder-only models contradicts \citet{icse_capture}, who concluded that the last two layers encode the syntactic relations better. \citet{icse_capture} uses a higher threshold (0.3) than in our work (0.05) and compares the heads with the best precision instead of those with the best F-score (our work). Our findings are consistent with the observations of \citet{earlybird}, who utilize early layers of CodeBERT for improved and efficient classification. 



Overall, in Figure \ref{fig: recall} we find that the models we studied encode 30-40\% of syntactic relations and around 50\% of data flow relations. This means that the majority of the code relations are still not encoded within the self-attention values. This raises the question - what relations are not encoded and how important are they for code understanding? To study the limitations quantitatively, we measure the similarity between model graphs and code graphs. The results are presented in Figure \ref{fig: similarity}. 

For all models, we find that the model graph has the highest similarity with DFG. However, smaller encoder-decoder models and deeper layers of larger encoder-decoder models have lower DFG similarity compared to encoder-only models. Thus, encoder-only models encode data flow relations better than encoder-decoder models and very large models encode data flow relations very poorly. 

When we study the syntax graphs in Figure \ref{fig: similarity}, we observe that model graphs of all models across each layer have much higher similarity with non-identifier graphs than with complete syntactic graphs. This means that the syntactic-identifier token relations are not encoded in the model graph. The reasoning is as follows. The edges in complete syntax graph comprises of all edges in non-identifier graph and additional syntactic-identifier edges. If these additional edges were present in the complete graph, the deletion cost and, hence, the overall cost for the complete syntax graph would have decreased. However, we observe a significant increase in cost per node, by a factor of 1.5-2. Thus, these additional edges relating syntactic and identifier tokens are not encoded in self-attention values, irrespective of model size and architecture. In fact, larger models encode syntactic relations poorly compared to smaller models.

\subsection{Analysis of Hidden Representation}
In our study of hidden representations using t-SNE, we find that the clustering of hidden representations does not follow syntactic relations in AST. In both the settings (hidden representation of tokens and distance matrix described in Section \ref{qualitative}) we find that the hidden representations create clusters based on token types rather than on syntactic relations. Due to space constraints, we show the t-SNE projections in Appendix \ref{tsne}.

In hidden representations (Figure \ref{fig: tsne_cb_5}), the clusters of syntactically related tokens such as, \texttt{def}, \texttt{(}, \texttt{)} and \texttt{:}, are not close to each other. But in distance matrix, certain syntactically related tokens do exist together. For the code in Figure \ref{fig: code}, we find that \texttt{def} is close to \texttt{(}, \texttt{)}, and \texttt{:} while \texttt{if} is close to \texttt{is} and \texttt{none} in the projection of fifth layer of CodeBERT (Figure \ref{fig: dist_cb_5}). Similarly, \texttt{not} and \texttt{in} occur together. However, identifiers are far from syntactic tokens including the token \texttt{=}, which usually establishes relations among variables. We found similar patterns for deeper layers of all models, while all tokens cluster together in the first few layers.

These observations contradict previous studies that use classifier and structural probing  \citep{troshin, probe_karmakar, probe_ahmed, icse_capture}. The previous works assume a linear encoding of information and hence, use a simple probe \citep{belinkov}. The studies conclude that hidden representations can encode syntactic relations among tokens. 

\begin{table}[tb]
\tiny
\begin{center}
\begin{tabular}{p{1.0cm}p{1.0cm}{c}p{0.3cm}p{0.3cm}p{0.3cm}p{0.3cm}p{0.3cm}}
\hline
\multicolumn{1}{c}{\bf Tokens}&\multicolumn{1}{c}{\bf Model}  &\multicolumn{1}{p{0.7cm}}{\bf No. of} &\multicolumn{5}{c}{\bf Label Accuracy}\\\cline{4-8}
& &\bf clusters & 2 & 3 & 4 & 5 & 6\\
 \hline
 & GraphCodeBERT& 9& 0.84& 0.78& 0.67& 0.67& 0.57\\
 \{Keyword- & CodeT5& 10& 0.83& 0.79& 0.70& 0.64& 0.60\\
 All\}& CodeT5+220M& 11& 0.78& 0.67& 0.58& 0.65& 0.58\\
 & CodeT5+220Mbi& 10& 0.64& 0.60& 0.52& 0.46& 0.44\\
 & CodeT5+770M& 9& 0.76& 0.70& 0.58& 0.61& 0.58\\
 & CodeRL& 13& 0.67& 0.67& 0.62& 0.67& 0.55\\
 & Codegen& 11& 0.61& 0.65& 0.56& 0.54& 0.48\\
 & CodeT5+2B& 9& 0.63& 0.66& 0.47& 0.55& 0.47\\
  \hline
 & GraphCodeBERT& 7& 0.79& 0.68& 0.52& 0.57& 0.49\\
 \{Keyword-& CodeT5& 6& 0.78& 0.66& 0.59& 0.55& 0.48\\
 Identifier\}& CodeT5+220M& 7& 0.82& 0.73& 0.65& 0.61& 0.52\\
 & CodeT5+220Mbi& 7& 0.65& 0.55& 0.51& 0.43& 0.41\\
 & CodeT5+770M& 5& 0.75& 0.69& 0.61& 0.59& 0.53\\
 & CodeRL& 5& 0.67& 0.63& 0.55& 0.53& 0.46\\
 & Codegen& 5& 0.68& 0.68& 0.54& 0.55& 0.60\\
 & CodeT5+2B& 5& 0.64& 0.63& 0.55& 0.42& 0.51\\
 \hline
\end{tabular}
\caption{The number of clusters formed by DirectProbe and label accuracy on hidden representation of last layer on distance prediction with 5 labels.}
\label{tab: dprobe_dist}
\end{center}
\end{table}

\begin{table}[tb]
\tiny
\begin{center}
\begin{tabular}{llccc}
\hline
\multicolumn{1}{p{0.4cm}}{\bf Tokens} &\multicolumn{1}{p{0.1cm}}{\bf Model}  &\multicolumn{1}{p{.7cm}}{\bf No. of}  &\multicolumn{2}{c}{\bf Label Accuracy}\\\cline{4-5}
 & &\bf clusters & Not Siblings & Siblings\\
 \hline
 & GraphCodeBERT& 4&0.76& 0.87\\
 \{Keyword-& CodeT5& 7& 0.82& 0.91\\
 All\} & CodeT5+220M& 3& 0.78& 0.94\\
& CodeT5+220Mbi& 6& 0.72& 0.78\\
& CodeT5+770M& 6& 0.81& 0.88\\
& CodeRL& 6& 0.79& 0.85\\
& Codegen& 4& 0.76& 0.85\\
& CodeT5+2B& 5& 0.48& 0.85\\
  \hline
 & GraphCodeBERT& 3& 0.75& 0.86\\
 \{Keyword-& CodeT5& 4& 0.80& 0.86\\
 Identifier\}& CodeT5+220M& 3& 0.80& 0.87\\
& CodeT5+220Mbi& 4& 0.58& 0.74\\
& CodeT5+770M& 4& 0.75& 0.87\\
& CodeRL& 4& 0.67& 0.78\\
& Codegen& 3& 0.77& 0.83\\
& CodeT5+2B& 3& 0.65& 0.76\\
 \hline
\end{tabular}
\caption{The number of clusters formed by DirectProbe and label accuracy on hidden representation of last layer on siblings prediction with 2 labels.}
\label{tab: dprobe_sib}
\end{center}
\end{table}

\begin{table}[tb]
\tiny
\begin{center}
\begin{tabular}{p{1.0cm}p{1.3cm}{c}p{0.4cm}p{0.6cm}p{0.6cm}}
\hline
\multicolumn{1}{c}{\bf Tokens}&\multicolumn{1}{c}{\bf Model}  &\multicolumn{1}{p{0.7cm}}{\bf No. of} &\multicolumn{2}{c}{\bf      Label Accuracy}\\\cline{4-6}
 & &\bf clusters & No Edge & Comes From & Computed From \\
 \hline
 & GraphCodeBERT& 7& 0.71& 0.94& 0.93\\
 \{Identifier-& CodeT5& 4& 0.57& 0.86& 0.90\\
 Identifier\} & CodeT5+220M& 4&0.69& 0.90& 0.88\\
 & CodeT5+220Mbi& 3&0.64& 0.84& 0.84\\
 & CodeT5+770M& 4&0.63& 0.89& 0.92\\
 & CodeRL& 6& 0.65& 0.85 & 0.84\\
 & Codegen& 5&0.63& 0.86& 0.92\\
 & CodeT5+2B& 4&0.63& 0.89& 0.92\\
  \hline
\end{tabular}
\caption{The number of clusters formed by DirectProbe and label accuracy on hidden representation of last layer data flow edge prediction with 3 labels.}
\label{tab: dprobe_dfg}
\end{center}
\end{table}

Using DirectProbe (see Appendix \ref{background}), we study both, what information is encoded in hidden representation and how - linearly or non-linearly. We report the number of clusters and clustering accuracy for the last layer in Tables \ref{tab: dprobe_dist}, \ref{tab: dprobe_sib} and \ref{tab: dprobe_dfg} (See Appendix \ref{clusters} for more layers and models). The number of clusters created by DirectProbe indicates whether the hidden representations encode a property linearly or non-linearly. Linear encoding results in the same number of clusters as the number of labels. 
For all three tasks, we observe a significantly higher number of clusters than labels across all models, usually twice as many. This means that hidden representations encode syntactic and data flow relations non-linearly. Thus, a simple probe is not sufficient to study hidden representation of cLLMs \citep{belinkov} 


In case of pre-trained models, we find that DirectProbe forms clusters with high accuracy on siblings and data flow tasks (Tables \ref{tab: dprobe_sib} and \ref{tab: dprobe_dfg}). But, on the tree distance tasks shown in Table \ref{tab: dprobe_dist}, the cluster accuracy is poor for $distacne > 2$ for \texttt{Keyword-All} token pairs and even poorer for \texttt{Keyword-Identifier} pairs. However for fine-tuned (CodeRL, CodeT5\_musu) and zero-shot (CodeT5+220Mbi, CodeT5+2B, CodeGen) models, the accuracy is poor on data flow and siblings task with \texttt{Keyword-Identifier} token pairs and dismal on distance prediction task.  

The observations imply that the hidden representations do not encode sufficient information for the distance prediction task. As described in Section \ref{probing}, this in turn implies that hidden representations of code models do not encode information about different identifier types and syntax structures. Surprisingly, the fine-tuned and zero-shot models additionally also do not properly understand which syntactic and identifier tokens are siblings and which tokens have data flow relations. 

\section{Discussion}
\subsection{Limitations of cLLMs} \label{limits}

Our analysis of attention maps reveals that they do not encode self-attention between syntactic and related identifier tokens. For example, in the best F-score case in Figure \ref{fig: explanation}, we observe that the keyword \texttt{if} pays attention to the related syntactic token \texttt{is}, but not to the related identifier \texttt{ignore}. 
The analysis of hidden representations reveals that they do not encode sufficient information to differentiate between common syntactic structures. 


We argue that these issues limit the ability of cLLMs to understand the program flow and what the code does. Program flow depends on the value of the expression associated with the conditional (\texttt{if, elif}) or loop (\texttt{for, while}). However, the syntactic tokens do not pay attention to the associated expression. Further, the hidden representations do not encode sufficient information to differentiate between the forms of expression. Thus, the model does not understand how to evaluate an expression - whether to use the value of the variable, evaluate a comparison or logical operator, or call a function. Due to the failure of models to understand the evaluation of the expression, they cannot reason about the execution path that will be taken. Given that a program can perform different operations depending on the execution path, the model cannot quite understand what the program does.

The evaluation of the expression, and thus the flow, may also depend on the input to the program. The input is usually not provided during training. However, even CodeRL, trained with feedback based on test cases, does not encode the information to understand the program flow. Further, these limitations exist irrespective of transformer architecture, size, or training objective. Thus, it could be a fundamental limitation of the transformer architecture on coding tasks.

\subsection{Code Property v/s Model Performance} \label{suggestion}
Models fine-tuned on a specific task perform better than pre-trained models on that task. However, the DirectProbe analysis reveals that pre-trained models encode syntactic information better than the fine-tuned models. Our findings are consistent with those of \citet{troshin}, whose classifier-based probing revealed that fine-tuned models encode syntactic information worse than pre-trained models. Our analysis additionally reveals that even pre-trained models do not encode syntactic-identifier relations necessary for understanding program flow. Further, \citet{code_summay} showed that models fine-tuned on summarization 
depend on shortcut cues such as function names and variables and not on code logic for correct summary.

Models with billions of parameters perform very well on code generation and in-filling tasks in a zero-shot manner. But our analysis reveals that they encode syntactic information very poorly. The repetitive nature of code corpora compared to natural language corpora \citep{naturalness, nl_pl_diff} results in memorization in cLLMs. However, multiple works have shown that larger models are more prone to memorizing training data compared to smaller models \citep{mem_rabin, mem_yang, large_fail}. Memorization, coupled with data contamination, results in good benchmark performance \citep{expl} but the benchmark performance do not translate to real-world performance \citep{hellendoorn, aye}.

\section{Conclusion}
In this paper, we critically examined arbitrary assumptions made in previous works on interpretability of cLLMs and demonstrated that these assumptions can lead to misleading conclusions. 

Further, with improved experimental setting, we conducted an in-depth analysis of self-attention and hidden representations of cLLMs. The analysis revealed that cLLMs
struggle to encode code relations between syntactic and identifier tokens. This 
restricts their ability to understand program flow and logic. 
We also observed that fine-tuned models and larger models with billions of parameters encode these relations poorly compared to smaller pre-trained models. It seems that fine-tuned and larger models rely on shortcut learning and memorized code 
instead of code understanding.

Our work 
contributes to designing more robust experiments to study interpretability of cLLMs. It also suggests that it is important to explore novel training techniques and/or architectures to enhance models' capability to encode code properties, instead of using larger models with memorization. In our future work, we aim to investigate more recent instruction-tuned models by extending this study to NL-PL alignment.

\section*{Limitations}
Broadly, our work has following limitations.

First, the models we analyzed use sub-word tokenizers but we performed analysis on code words. For the code word level analysis, we merged the sub-words and the attention values / hidden representations of the corresponding sub-words by taking the mean of the values. While this is a standard practice in the analysis of attention maps and hidden representation, it can also introduce minor discrepancies in the results. 

Second, we only study the cases where codes are input. Thus the tasks involving text-to-code are not analyzed in our work. It is also not trivial to extend our work to text-to-code setting. Code models and LLMs in general are highly sensitive to minor changes in input. Due to this sensitivity, semantically similar texts can lead to significantly different output. We aim to extend this work to text-to-code settings by creating a statistical method to analyze NL-PL alignment in future work. Despite this limitation, our work has relevance for code-to-code and code-to-text applications.

Third, our work focuses on Python code, despite some of the models being trained on other programming languages (PLs) along with Python. Our work focuses on Python, because (1) the performance of cLLMs is much better on Python compared to other PLs and (2) Python has become the primary focus of many recent works and most recently released models have checkpoints specifically fine-tuned for Python code. However, limiting the analysis to Python also prevents us from studying certain programming constructs, such as type systems and cLLMs' understanding of types. 

\section*{Acknowledgements}
This research work was supported by the National Research Center for Applied Cybersecurity ATHENE.

\bibliography{custom}

\appendix

\section{Hardware Details}
We first perform a forward pass through the models on an Nvidia A6000 48GB GPU and store the attention and hidden representation for experiments. All experiments are then run on an AMD Ryzen Threadripper 5975WX with 32 cores. 

\section{Background} \label{background}
\begin{figure*}
    \centering
    \begin{subfigure}[b]{0.4\textwidth}
    \begin{center}
     \includegraphics[width=\linewidth]{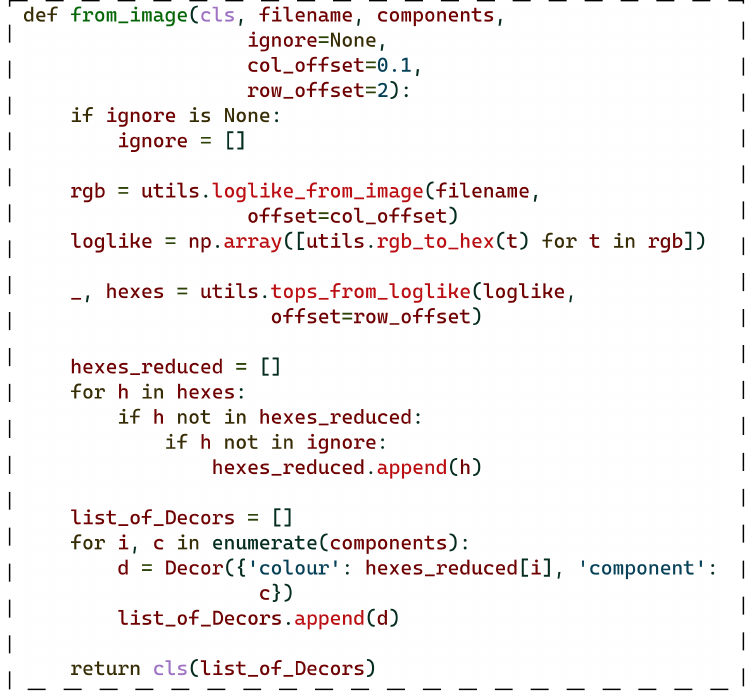}   
    \end{center}
    \caption{}
    \label{fig: code}
    \end{subfigure}
    \hfill
    \begin{subfigure}[b]{0.4\textwidth}
    \begin{center}
    \includegraphics[width=\linewidth]{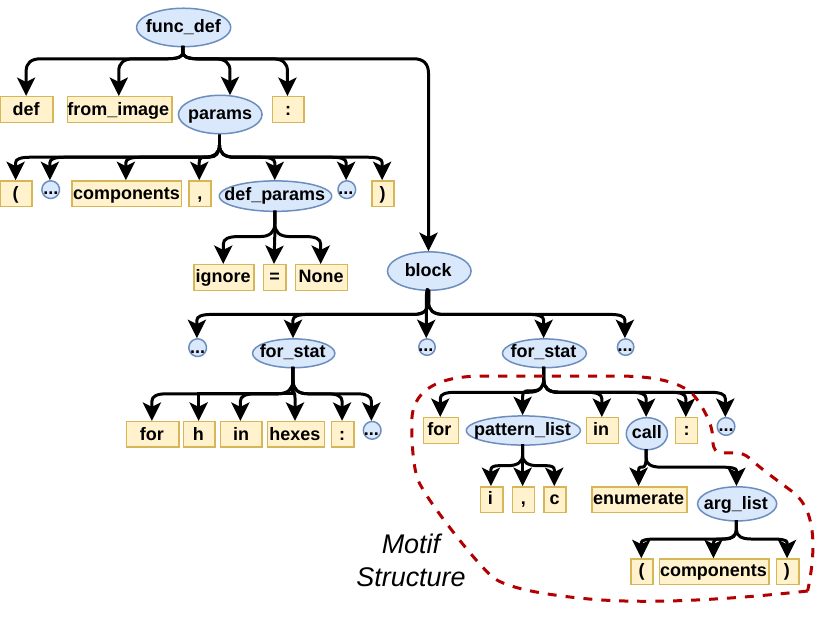}
    \end{center}
    \caption{}
    \label{fig: ast}
    \end{subfigure}
    \hfill
    \begin{subfigure}[b]{0.4\textwidth}
    \begin{center}
    \includegraphics[width=\linewidth]{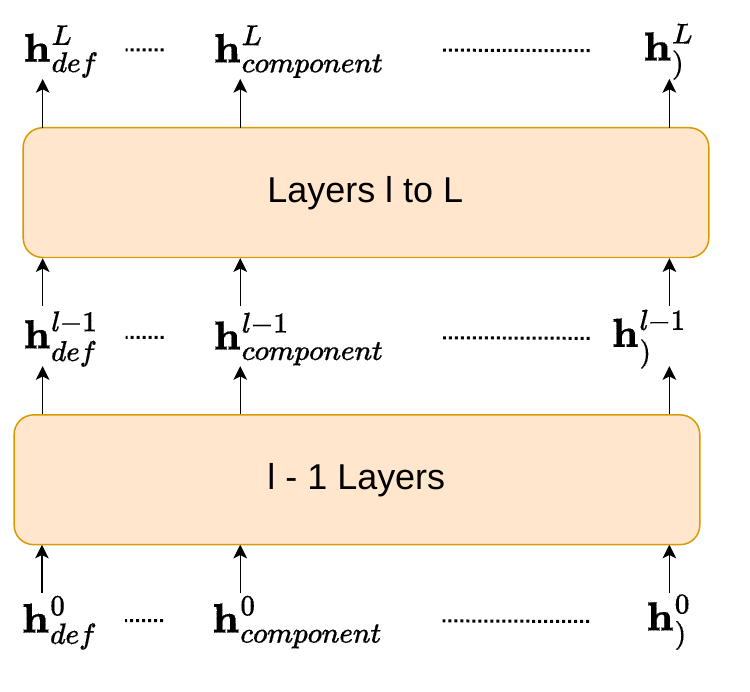}
    \end{center}
    \caption{}
    \label{fig: hidden_repr}
    \end{subfigure}
    \hfill
        \begin{subfigure}[b]{0.4\textwidth}
    \begin{center}
    \includegraphics[width=\linewidth]{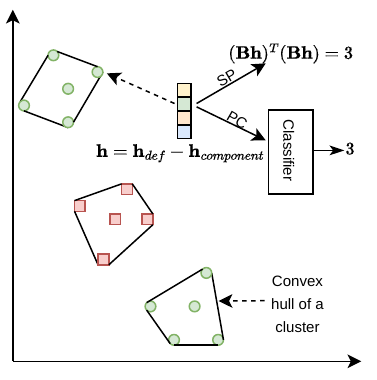}
    \end{center}
    \caption{}
    \label{fig: probe}
    \end{subfigure}
\caption{A python code snippet \textbf{(a)} and it's (partial) AST \textbf{(b) showing an example of motif structure}; Illustration of hidden representation in a transformer model \textbf{(c)}; An illustration of structural probe (SP), probing classifier (PC) and convex hull created by DirectProbe \textbf{(d)}.}
\end{figure*}
\textbf{Attention Analysis.} In NLP, attention analysis investigates whether self-attention corresponds to linguistic relations among input tokens. For cLLMs, attention analysis quantifies how well self-attention encodes relations among code tokens, such as relations in an AST.

\textbf{Probing on Hidden Representation} is a technique to study the properties encoded in the hidden representations \citep{belinkov}. Due to the many limitations of classifier or structural probe based probing techniques \citep{control_task, nlsp, tale}, we use DirectProbe \citep{directprobe}, a non-classifier-based probing technique. DirectProbe clusters the hidden representations of a specific layer based on labels for the property we want to study. Then, the convex hull of these clusters (Figure \ref{fig: probe}) can be used to study how well hidden representations encode information about that property. The basic idea is that a good-quality representation will have well-separated clusters, while linear encoding of a property will result in each label having one cluster. The quality of clustering can be evaluated by predicting clusters for a hold-out test set.

\textbf{Abstract Syntax Trees (ASTs)}
are data structures that represent the syntactic structure of a code. The leaf nodes of the tree represent code tokens, and internal nodes represent different constructs of the code such as \texttt{if-else} block, \texttt{identifiers}, or \texttt{parameters}. A partial AST\footnote{We use tree-sitter (\href{https://tree-sitter.github.io/tree-sitter/}{https://tree-sitter.github.io/tree-sitter/})to obtain AST of a code.} for a Python code snippet is shown in Figure \ref{fig: ast}.

\textbf{Data Flow Graphs (DFGs)}
have nodes representing variables and edges depicting how the values flow from one variable to another. We adopt the approach by \citet{graphcodebert} to obtain the data flow relations, 
with two types of data flow relations, viz. \texttt{ComesFrom} and \texttt{ComputedFrom}.

\textbf{Motif Structure} \citet{icse_capture} defines motif structure as a non-leaf node in the AST with all it's children and assume there is a syntactical relation between all leaf nodes (i.e. code tokens) of a motif structure. We show motif structure in Figure \ref{fig: ast}.

\textbf{Transformer and Self-attention.}
A Transformer model consists of $L$ stacked transformer blocks. The core mechanism of a transformer block is self-attention. Given a code $c = \{c_1, c_2, ..., c_n\}$ of length $n$, the self-attention mechanism assigns an input token $c_i$ attention values over all input tokens. The code $c$ is first transformed into a list of $d$-dimensional vectors $\mH^0 = [\vh_1^0, \vh_2^0, ..., \vh_n^0]$. The transformer model transforms $\mH^0$  into a new list of vectors $\mH^L$. A layer $l$ takes the output of the previous layer $\mH^{l-1}$ as input and computes $\mH^l= [\vh_1^l, \vh_2^l, ..., \vh_n^l]$. $\vh_i^l$ is the hidden representation of $i^{th}$ word at layer $l$, as shown in Figure \ref{fig: hidden_repr}. Attention values for layer $l$ are computed as
 \begin{equation}
    Attention (\mQ, \mK, \mV) = softmax(\frac{\mQ\mK^T}{\sqrt d})\mV
\end{equation}
where $\mQ = \mH^{l-1}\mW_Q^l$, $\mK = \mH^{l-1}\mW_K^l$ and $\mV = \mH^{l-1}\mW_V^l$. In practice, a layer $l$ contains multiple heads, each with its own $\mW_Q^l$, $\mW_K^l$, $\mW_V^l$ matrices. Each head thus has a set of attention values among each pair of input tokens, which constitute the attention map for that head (Figure \ref{fig: explanation}).

\begin{table*}[htb]
\begin{center}
\begin{tabular}{lccccc}
\hline
\multicolumn{1}{c}{\bf Range}&\multicolumn{1}{c}{\bf CodeBERT}  &\multicolumn{1}{c}{\bf GraphCodeBERT}  &\multicolumn{1}{c}{\bf UniXcoder} &\multicolumn{1}{c}{\bf CodeT5}&\multicolumn{1}{c}{\bf PLBART}\\
 \hline
 0.0& 59.13& 70.3& 67.28& 51.92& 74.63\\
 0.0 - 0.05& 39.25& 28.58& 31.88& 46.23& 74.27\\
 0.05 - 0.3& 1.48& 1.00& 0.76& 1.64& 0.97\\
 above 0.3& 0.14& 0.12& 0.08& 0.22& 0.13\\
 \hline
\end{tabular}
\caption{Percentage of attention values in differenr range.}
\label{tab: distr}
\end{center}
\end{table*}
\section{Model Details} \label{model}
We ran our experiments on multiple openly-available models chosen to represent different model architectures, sizes, training objectives and trained on different dataset. The models have parameters ranging from 110M to 3.7B parameters. We perform the experiments with pre-trained and fine-tuned models as well as models which show good benchmark performance in zero-shot setting. Among the pre-trained models, we consider CodeBERT \citep{codebert}, GraphCodeBERT \citep{graphcodebert}, UniXcoder \citep{unixcoder}, CodeT5 \citep{codet5} and PLBART \citep{plbart}, CodeT5+220M \citep{codet5p} and CodeGen \citep{codegen}.

\textbf{CodeBERT} is an encoder-only bi-directional transformer with 220M parameters comprising of 12 layers, each layer having 12 heads. It has been trained on CodeSearchNet (CSN) \citep{csn} dataset with two pre-trained objectives. Masked Language Modeling (MLM) objective is used with bimodal (NL-PL pair) data, the model is trained with and Replaced Token Detection (RTD) with unimodal (only PL) data. 

\textbf{GraphCodeBERT} uses the same architecture as CodeBERT but also takes nodes of the data flow graph (DFG) of the code as inputs with special position embeddings to indicate which tokens are nodes of DFG. It is also trained on CSN dataset. The model is first trained with MLM objective, followed by edge prediction in data flow graph and node alignment between code tokens and DFG nodes.

\textbf{UniXcoder} is an encoder-decoder model with 220M parameters. However, the model can be used in encoder-only, decoder-only or encoder-decoder mode using a special input token, \texttt{[MODE]}. It is also trained on CSN dataset and taked flattened ASTs of code as part of it's input during training. The model is trained with masked spans prediction, masked language modeling, multi-modal contrastive learning, whereby positive pairs are created using dropout, and cross-modal generation.

\textbf{CodeT5} is an encoder-decoder model with 220M parameters trained on CSN dataset with identifier-aware and bimodal-dual generation objective. Identifier-aware pretraining uses masked span prediction, identifier tagging and masked identifier prediction alternatively to make the model attend to identifiers while bimodal-dual generation consists of NL to PL generation and PL to NL generation. Along with pre-trained CodeT5, we also experiment with CodeT5 fine-tuned for summarization task. Further, we include a larger CodeT5 model trained with next token prediction task and then trained on Python code and CodeRL \citep{coderl} which is fined-tuned for code generation in an actor-critic setup with feddback from test cases .    

\textbf{PLBART} PLBART is an encoder-decoder model with 110M parameters comprising of 6 encoder layers, each with 12 heads. The model is trained with 3 denoising objectives - token masking, token deletion and token infilling - on NL and PL data from Google BigQuery       \footnote{\href{https://console.cloud.google.com/marketplace/details/github/github-repos}{https://console.cloud.google.com/marketplace/github-repos}}.

\textbf{CodeT5+} is a family of models trained with span denoising, causal LM, contrastive loss and matching loss. We experiment with the 220M, 770M and 2B variants of CodeT5+ model. The 220M and 770M have the same architecture as CodeT5, while the 2B variant follows the architecture of CodeGen-mono 350M for encoder and CodeGen-mono 2B for decoder. CodeT5+ can be used in encoder-only, encoder-decoder and decoder-only setup. So we also study the decoder of the 2B variant. Further, we also study the 220M-bimodal variant which can be used for code summarization and retrieval in zero-shot manner.  

\textbf{CodeGen} CodeGen is a decoder-only model trained with fill-in-the-middle objective \citep{fim} to provide bi-directional context during training. We experiment with the 3.7B variant of the model with 16 layers and 16 heads in each layer. The model can be used for code generation in zero-shot setting.

\section{Datset Details}
\label{dataset}
CodeSearchNet \citep{csn} dataset consists of 2 million comment-code pairs from 6 programming languages and is a commonly used dataset to pre-train models.  The programming languages are Go, Java, JavaScript, PHP, Python and Ruby. The codes in the dataset are scrapped from GitHub and filtered to only contain codes with permissible licenses. Different codes have different licenses and the details of those licenses is available in the dataset. We experiment with the Python codes from test split of CSN \citep{csn}. 

We chose CSN for our experiments because most of the models we considered have been pre-trained on CSN or CSN augmented with additional data. Due to this, the effect of data distribution shift is minimized.

Before performing analysis we pre-process the dataset by removing any docString and code comments from the dataset. CodeBERT, GraphCodeBERT and UniXcoder has a maximum input token length of 512 tokens. So, we create a subset consisting of codes with less than 500 tokens post tokenization. CSN consists a list of code tokens for each token. For merging attention and hidden representation of sub-tokens, we use this list to keep track of where a token has been split by tokenizer. However, the list splits \texttt{*args} into \texttt{*} and \texttt{args} and \texttt{**kwargs} into \texttt{*}, \texttt{*} and \texttt{kwargs}. In Python, \texttt{*} is used for iterator unpacking and \texttt{**} for dictionary unpacking. So, to differentiate the two, we merger the \texttt{*}s of \texttt{kwargs}. From he pre-processed dataset, we randomly sample 3000 python code and run our experiments on these codes.

\section{Attention Distribution} \label{distrib}
In Table \ref{tab: distr}, we present the percentage of attention values which are 0, between 0 - 0.05, between 0.05 - 0.3 and more than 0.3. Note that we assume any value below 0.001 to be 0.

\section{Additional Attention Analysis Results}
\begin{figure*}[htbp]
    \centering
    \begin{subfigure}[b]{0.99\textwidth}
     \includegraphics[height=0.165\linewidth]{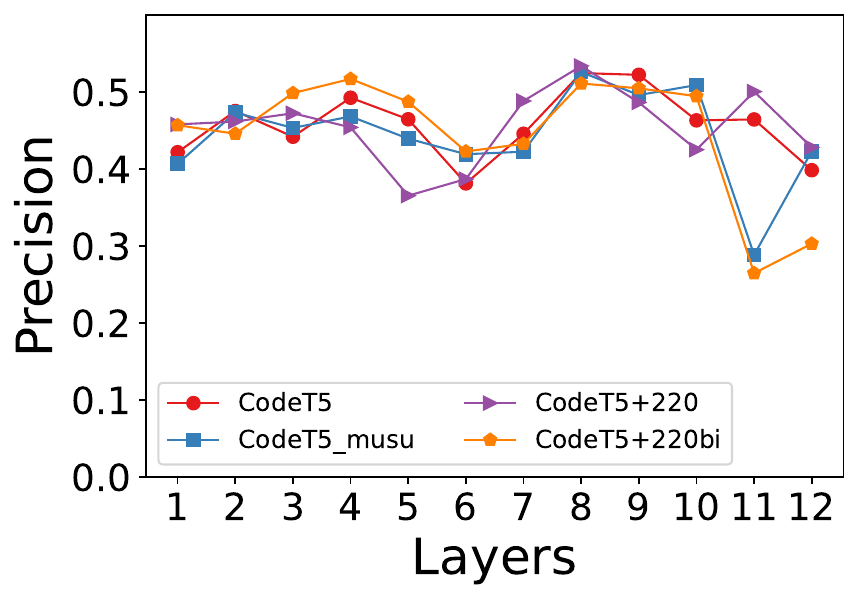} 
     \includegraphics[height=0.165\linewidth]{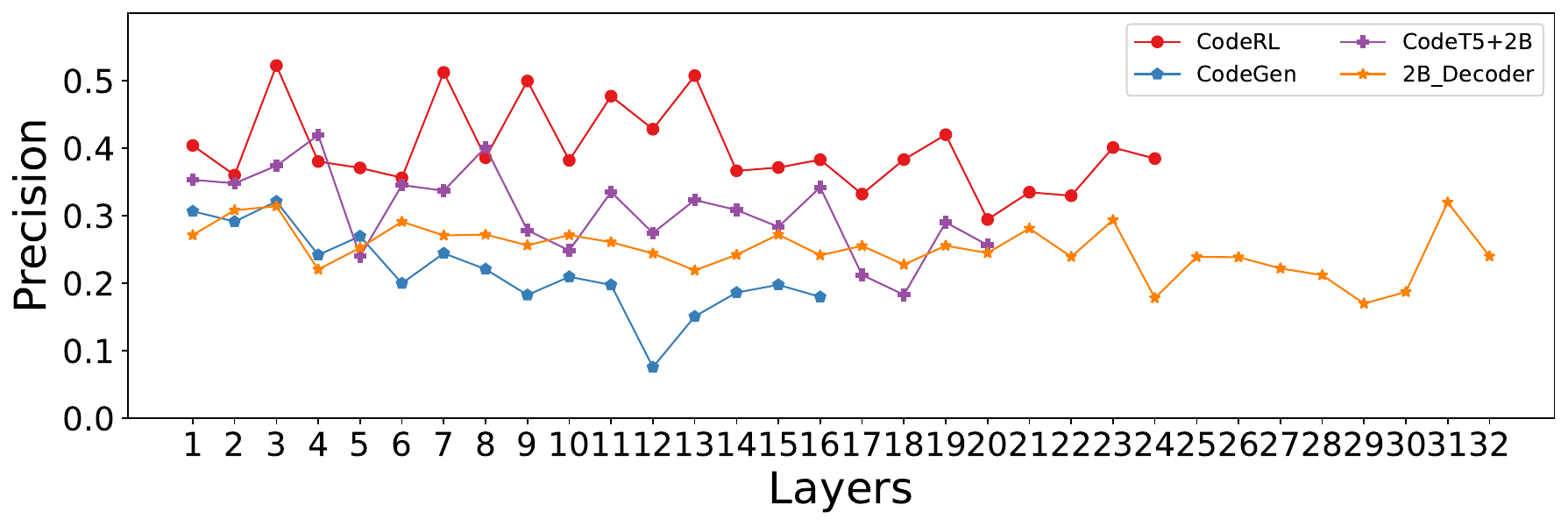} 
     \includegraphics[height=0.165\linewidth]{results_/recallandprecision/pre_ast_fine.pdf} 
    \end{subfigure}
        \begin{subfigure}[b]{0.99\textwidth}
     \includegraphics[height=0.166\linewidth]{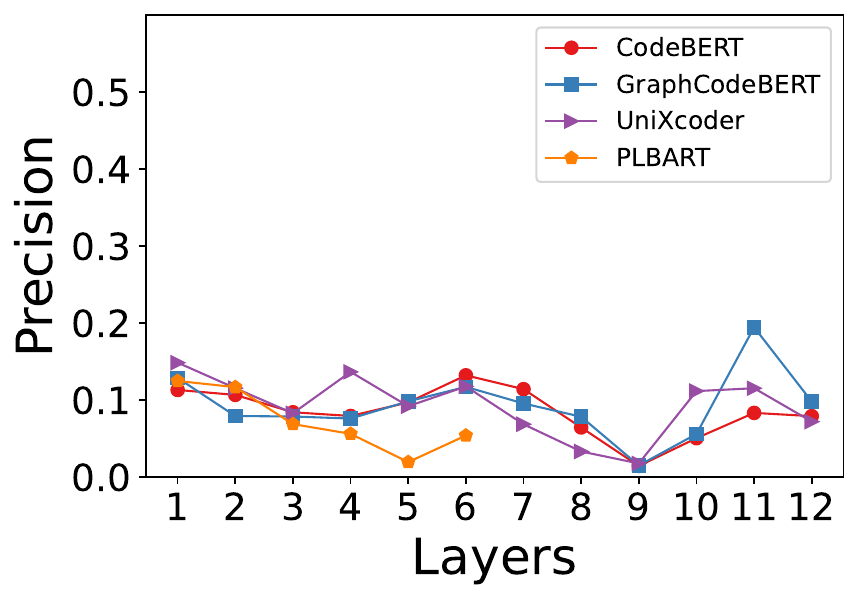} 
     \includegraphics[height=0.166\linewidth]{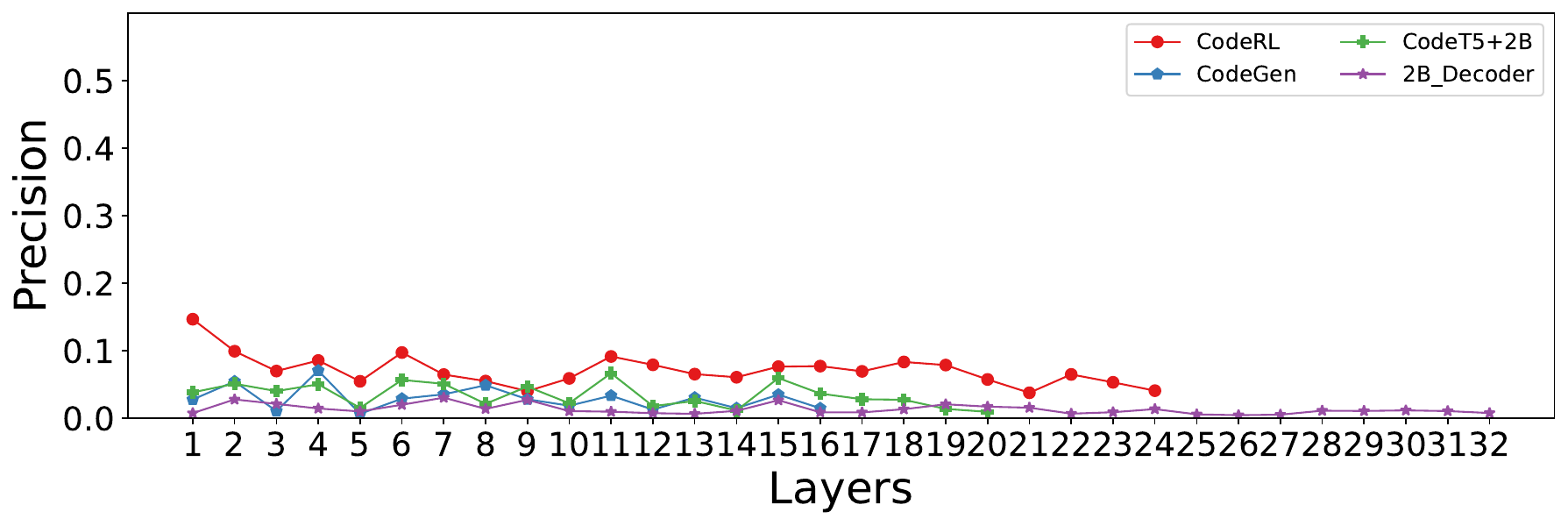} 
     \includegraphics[height=0.166\linewidth]{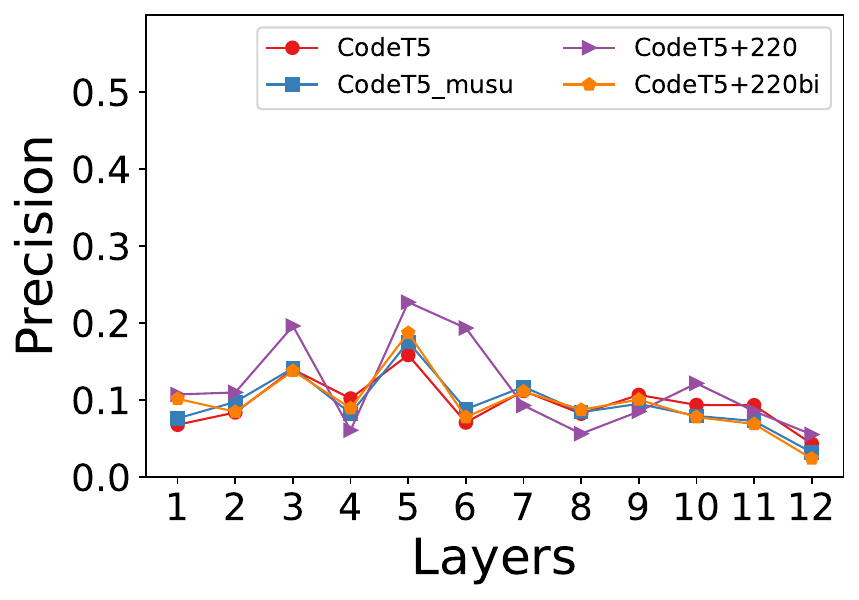} 
    \end{subfigure}
    \caption{Precision of model graphs with syntax graphs (top) and data flow graphs (bottom).}
    \label{fig: precision}
\end{figure*}

\begin{figure*}[htbp]
\centering
    \begin{subfigure}[b]{0.99\textwidth}
     \includegraphics[height=0.35\linewidth]{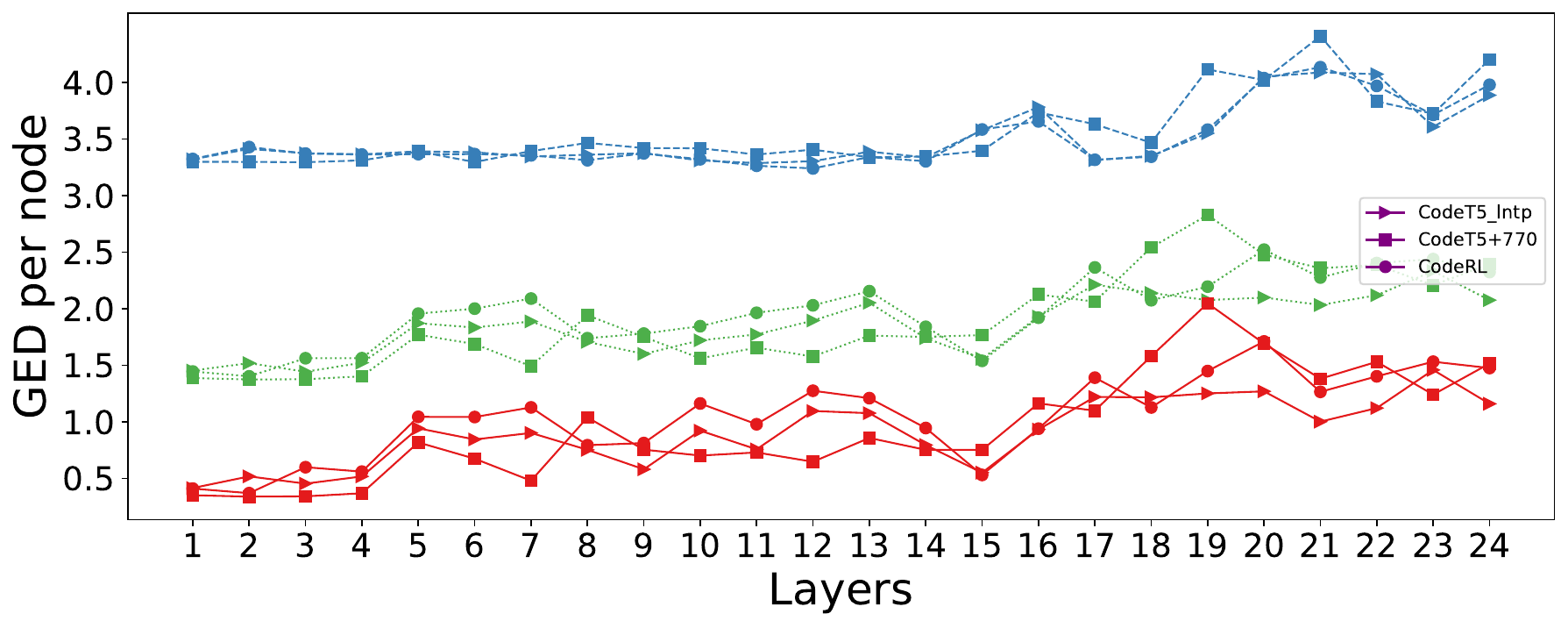} 
    \end{subfigure}
    \begin{subfigure}[b]{0.99\textwidth}
    \centering
    \includegraphics[width=0.478\textwidth]{results_/recallandprecision/legend.pdf} 
    \end{subfigure}
    \caption{Graph edit distance (GED) per node (lower value show higher similarity) of model graph from DFG, non-identifier syntax graph and complete syntax graph for various models. }
    \label{fig: similarity_other}
\end{figure*}

We present some additional results for attention analysis such as precision of model graphs (Figure \ref{fig: precision}) with syntax graphs and data flow graphs and graph edit distance (Figure \ref{fig: similarity_other})for some more models.

\begin{figure*}[htbp]
    \centering
    \begin{subfigure}[b]{0.99\textwidth}
     \includegraphics[height=0.18\linewidth]{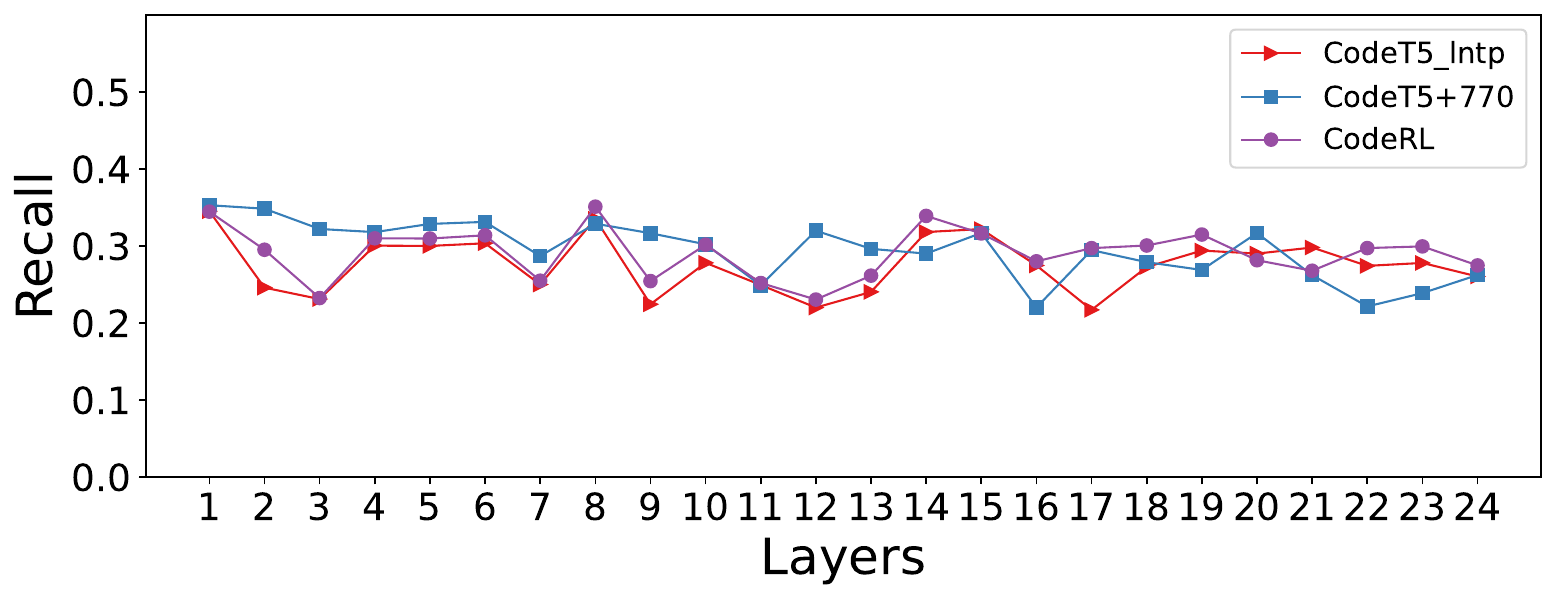} 
     \includegraphics[height=0.18\linewidth]{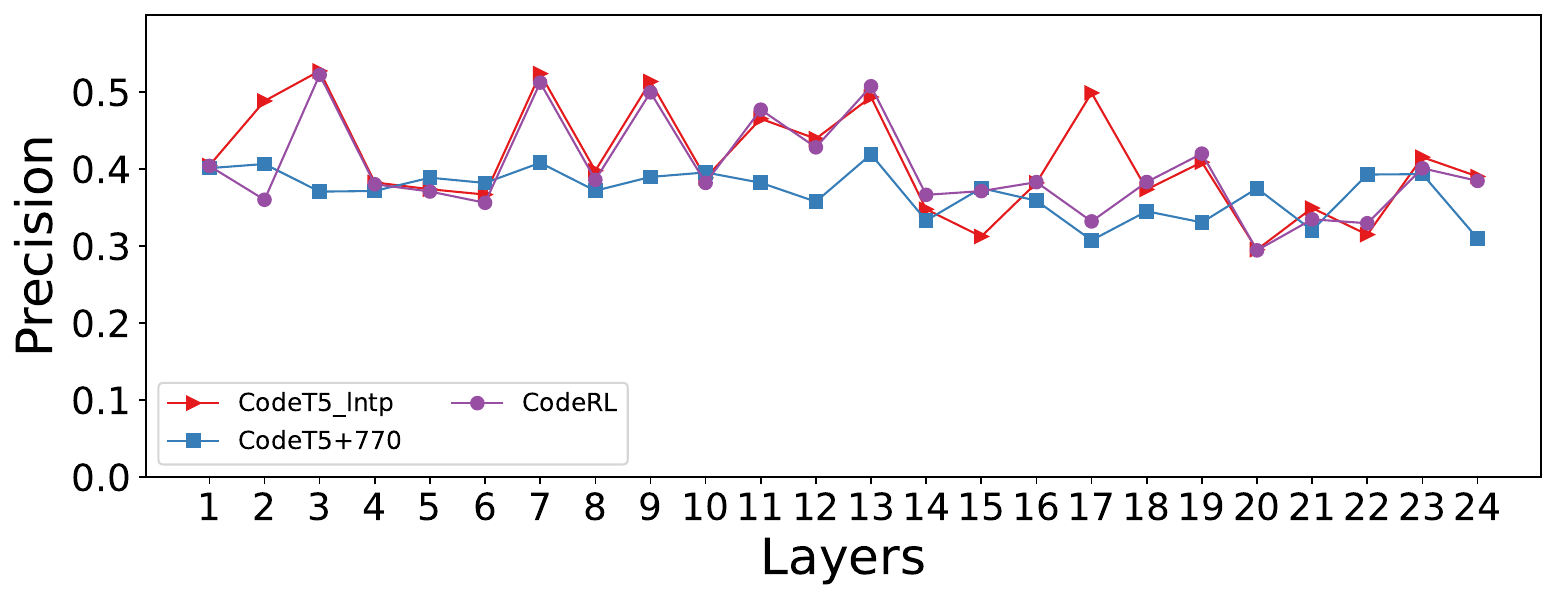} 
    \end{subfigure}
        \begin{subfigure}[b]{0.99\textwidth}
     \includegraphics[height=0.18\linewidth]{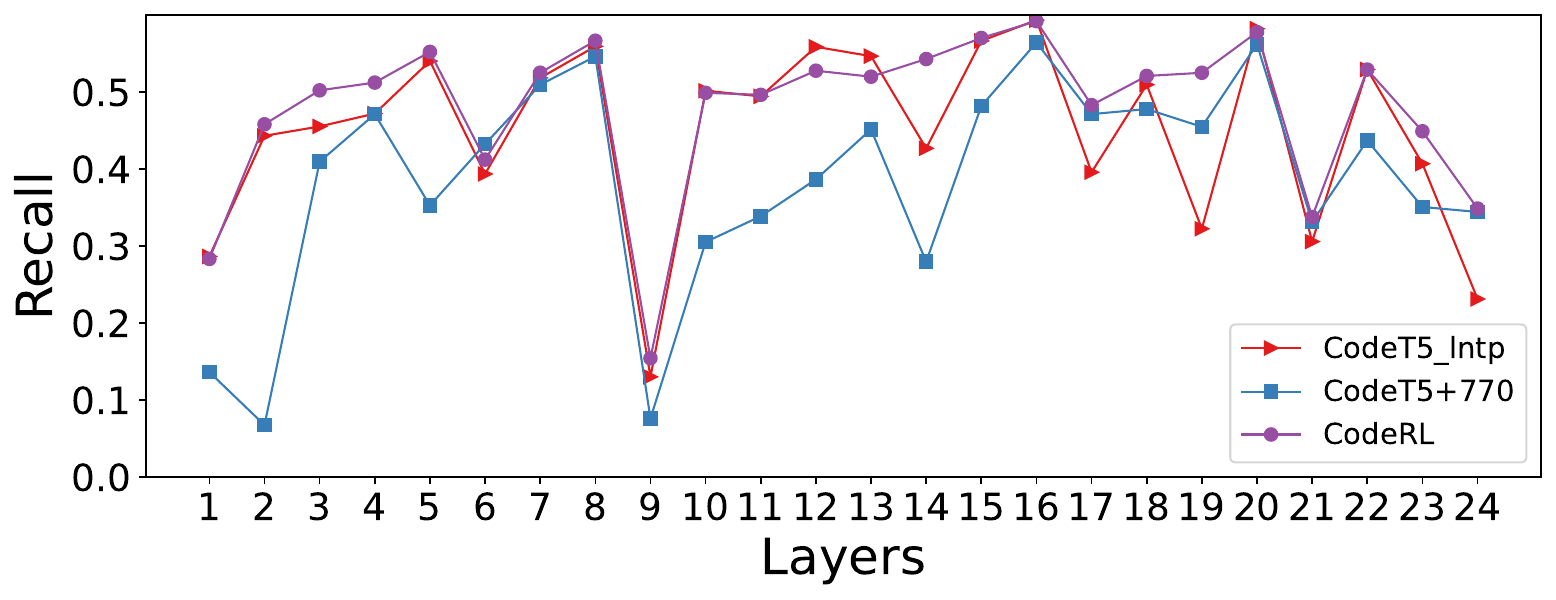} 
     \includegraphics[height=0.18\linewidth]{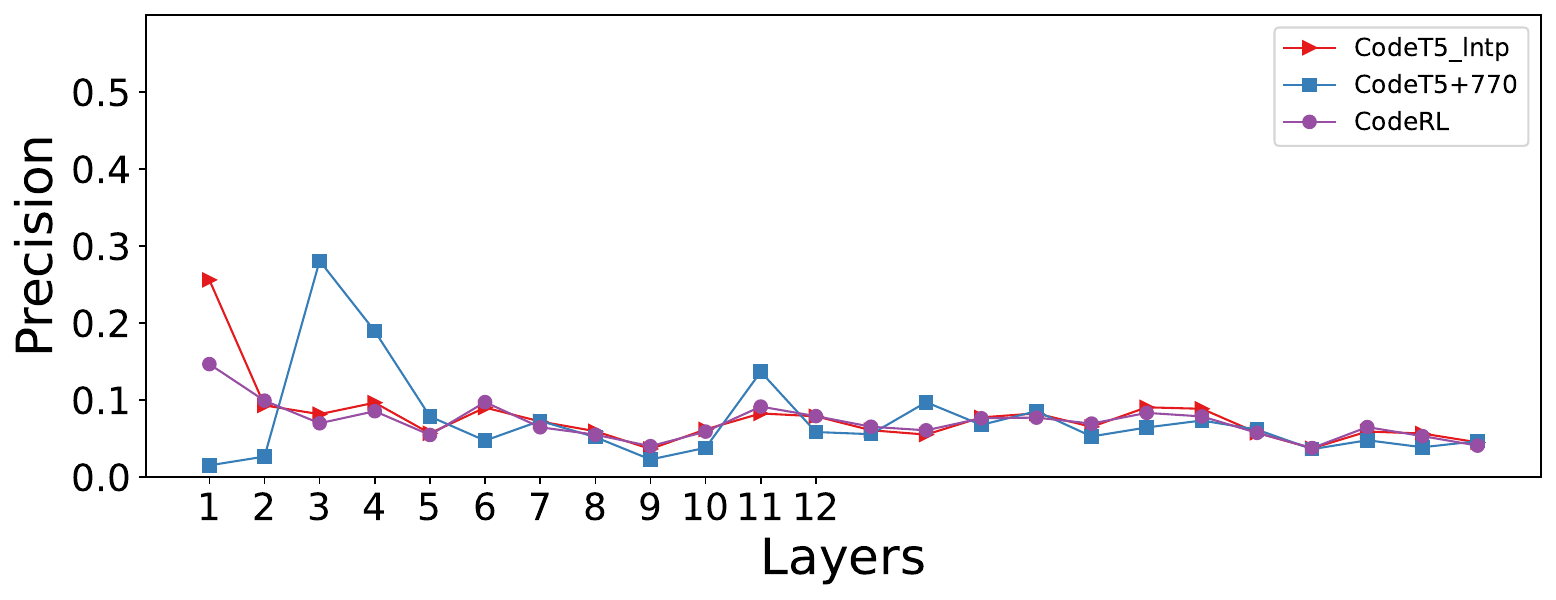} 
    \end{subfigure}
    \caption{Precision and Recall of model graphs with syntax graphs (top) and data flow graphs (bottom).}
    \label{fig: recallandprecision}
\end{figure*}

\section{t-SNE}
\label{tsne}
\begin{figure*}[h]
\begin{center}
\includegraphics{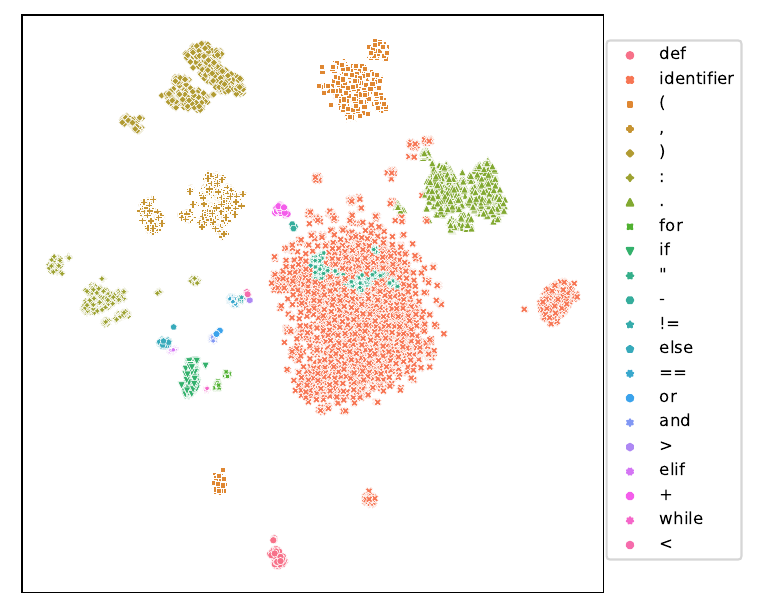}
\end{center}
\caption{t-SNE visualization of hidden representation of layer 5 of CodeBERT for selected token types.}
\label{fig: tsne_cb_5}
\end{figure*}

We select 100 codes with at least 100 code tokens and get the hidden representation for each token. We then select hidden representation of the token types shown in Figure \ref{fig: tsne_cb_5}. We ran t-SNE on the selected hidden representation with different perplexity value \citep{tsne} from 5 to 50 for all layers of all models. Increasing the perplexity value only made the clusters tighter but the overall distribution of points remained similar. So, the conclusion is not affected by perplexity value. We set the number of iterations to 50K, ensuring t-SNE always converges (no change in error for at least 300 iterations). We found that for all layers, tokens of same type were closer, though the clustering of same token types became tighter for deeper layers. We show the visualization for fifth layer of CodeBERT with perplexity of 50 in Figure \ref{fig: tsne_cb_5}.

\begin{figure*}[h]
\begin{center}
\includegraphics[width=0.96\textwidth]{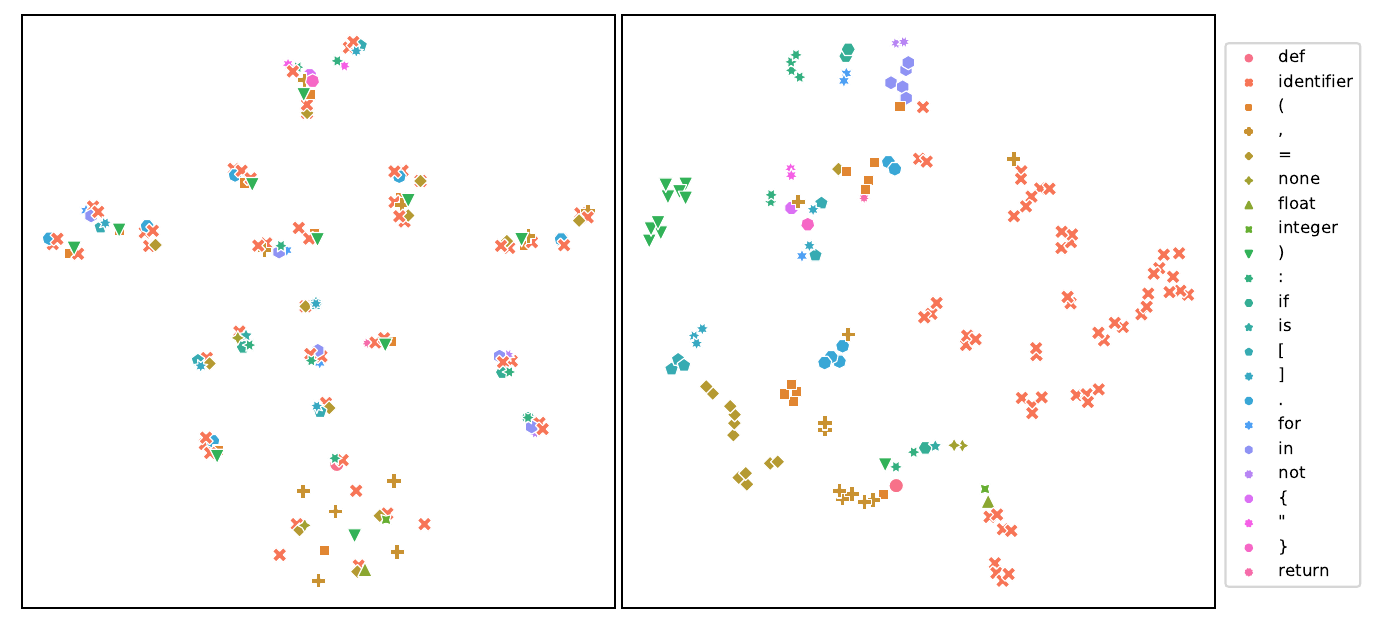}
\end{center}
\caption{t-SNE visualization of distance matrix for AST(left) and hidden representation (right) of layer 5 of CodeBERT for code in Figure \ref{fig: code}.}
\label{fig: dist_cb_5}
\end{figure*}

We create a distance matrix for both the tree distance in AST and distance between hidden representation of tokens for a few code. We run t-SNE till convergence with perplexity values 5 and 10 and found the distribution to be similar. We again observed clusters of tokens of same types for hidden representation, unlike clusters of AST distance matrix. The clusters are closer for earlier layers and farther for deeper layers. We show the visualization for fifth layer of CodeBERT for code in Figure \ref{fig: code} in Figure \ref{fig: dist_cb_5}. 

We use the t-SNE implementation provided by the sci-kit learn library\footnote{\href{https://scikit-learn.org/stable/modules/generated/sklearn.manifold.TSNE.html}{https://scikit-learn.org/generated/sklearn.TSNE.html}}.

\section{DirectProbe Experiment Details}
\label{directprobe}
For siblings and tree distance prediction tasks, the first token is of one of the following token types: \texttt{def for if none else false true or and return not elif with try raise except break while assert print continue class}.

For distance prediction task, we randomly sample 160 codes. We select the code pairs at a maximum distance of $6$, ensuring  first token is of one of the selected tokens types. The second token can be of any type. We then select 1300 code pairs for each layer resulting in a dataset of 6500 data points. We split it into train and test set in the ration of 80:20. We follow the same steps for \texttt{Keyword-Identifier} too, with the difference that we use 450 codes and the second token is of type \texttt{identifier}.

For distance prediction task, we randomly sample 100 codes. We first select all tokens which are one of the selected token types. We then select equal number of siblings and non-siblings for each of these selected tokens. From this, we randomly sample 1500 siblings and 1500 non-siblings resulting in 3000 data points. We split it into train and test set in the ration of 80:20. We follow the same steps for \texttt{Keyword-Identifier} too, with the difference that we use 300 codes and the second token is of type \texttt{identifier}.

For data flow edge prediction task, we randomly sample 130 codes. We first select an identifier and then the tokens which has a data flow edge with the first token. We then select $n$ tokens which do not have data flow edge with the first token, where, 
\begin{equation}
    \small n = \frac{\max(num(ComesFrom), num(ComputedFrom))}{2}.
\end{equation} From the selected pairs, we randomly sample 1500 pairs for each label resulting in 4500 data points. We split it into train and test set in the ration of 80:20. 

In all tasks, we ensure that the same data points are used for all models and layers.

\section{DirectProbe Results and Cluster Statistics}
\label{clusters}
In this section, we provide the statistics of size and label of cluster created by DirectProbe for last layer of some of the models and the results of experiments with DirectProbe for middle and layers of some models and last layers of models not reported in the main text. Analysis with DirectProbe is presented in Tables \ref{tab: dprobe_dist_ap}, \ref{tab: dprobe_sib_ap} and \ref{tab: dprobe_dfg_ap}. The cluster statistics are presented in Tables \ref{tab: cl_dist}, \ref{tab: cl_sib} and \ref{tab: cl_dfg}.

\begin{table*}[htb]
\tiny
\begin{center}
\begin{tabular}{llcccccccc}
\hline
\multicolumn{1}{c}{\bf Tokens}&\multicolumn{1}{c}{\bf Model (Layer)}  &\multicolumn{1}{p{1.2cm}}{\bf No. of clusters}  &\multicolumn{2}{c}{\bf Distance } &\multicolumn{5}{c}{\bf Label Accuracy}\\\cline{4-5} \cline{6-10}
& & &  Min & Avg & 2 & 3 & 4 & 5 & 6\\
 \hline
 &CodeBERT (5) &  9& 0.0& 1.09& 0.87& 0.85& 0.74& 0.72& 0.62\\
 & CodeBERT (9) & 9& 0.0& 1.36& 0.89& 0.81& 0.72& 0.72& 0.61\\
 & CodeBERT (12) & 10& 0.0& 1.27& 0.85& 0.75& 0.73& 0.68& 0.55\\
 &GraphCodeBERT (5) & 11& 0.0& 3.99& 0.88& 0.84& 0.75& 0.70& 0.63\\
 & GraphCodeBERT (9) & 9& 0.0& 1.74& 0.83& 0.81& 0.69& 0.68& 0.62\\
 \{Keyword-All\}& UniXcoder (5) & 10& 0.0& 1.87& 0.86& 0.82& 0.72& 0.71& 0.66\\
 & UniXcoder (9) & 9& 0.0& 0.70& 0.77& 0.77& 0.69& 0.63& 0.63\\
 & UniXcoder (12) & 13& 0.0& 2.59& 0.41& 0.55& 0.42& 0.48& 0.51\\
 &CodeT5 (5) & 9& 0.0& 1.65& 0.79& 0.80& 0.70& 0.67& 0.65\\
 & CodeT5 (9) & 13& 0.0& 8.50& 0.85& 0.83& 0.64& 0.70& 0.67\\
 & PLBART (3) & 13& 0.0& 2.60& 0.79& 0.77& 0.62& 0.70& 0.57\\
 & PLBART (6) & 9& 0.0& 1.88& 0.83& 0.83& 0.77& 0.70& 0.60\\
 &CodeT5+220M (5) & 13& 0.0& 0.49& 0.80& 0.74& 0.61& 0.65& 0.58\\
 &CodeT5220Mbi (5) & 15& 0.0& 1.70& 0.81& 0.70& 0.54& 0.55& 0.61\\
 &CodeT5770M (12) & 11& 0.0& 1.06& 0.76& 0.76& 0.68& 0.62& 0.59\\
  &CodeRL (12) & 13& 0.0& 1.59& 0.78& 0.72& 0.61& 0.64& 0.55\\
 &CodeT5\_{musu} (5) & 13& 0.0& 3.38& 0.76& 0.72& 0.57& 0.66& 0.59\\
 &CodeT5\_{musu} (12) & 11& 0.0& 1.51& 0.75& 0.70& 0.53& 0.56& 0.57\\
&CodeT5\_{lntp} (12) & 14& 0.0& 3.12& 0.79& 0.72& 0.60& 0.65& 0.55\\ 
&CodeT5\_{lntp} (24) & 10& 0.0& 0.85& 0.76& 0.72& 0.52& 0.64& 0.57\\
&Codegen (8) & 12& 0.0& 87.01& 0.73& 0.73& 0.59& 0.68& 0.48\\
&CodeT5+2B (10) & 10& 0.0& 8.26& 0.73& 0.74& 0.63& 0.65& 0.56\\
&CodeT5+2B\_dec (16) & 9& 0.0& 5.00& 0.58& 0.62& 0.45& 0.48& 0.40\\
&CodeT5+2B\_dec (32) & 12& 0.0& 12.90& 0.5& 0.56& 0.45& 0.44& 0.40\\

  \hline
 &CodeBERT (5) &  5& 0.0& 0.06& 0.86& 0.74& 0.64& 0.68& 0.59\\
 & CodeBERT (9) & 7& 0.0& 3.41& 0.89& 0.77& 0.63& 0.65& 0.57\\
 & CodeBERT (12) & 7& 0.0& 0.53& 0.82& 0.66& 0.56& 0.53& 0.51\\
 &GraphCodeBERT (5) & 5& 0.0& 0.05& 0.83& 0.70& 0.63& 0.64& 0.56\\
 & GraphCodeBERT (9) & 7& 0.0& 2.79& 0.83& 0.69& 0.60& 0.62& 0.56\\
 \{Keyword-Identifier\}& UniXcoder (5) & 7& 0.0& 2.33& 0.82& 0.66& 0.61& 0.61& 0.49\\
  &UniXcoder (9) & 7& 0.0& 5.07& 0.69& 0.61& 0.53& 0.55& 0.44\\
 &UniXcoder (12) & 9& 0.0& 5.37& 0.37& 0.49& 0.36& 0.32& 0.34\\
 &CodeT5 (5) & 7& 0.0& 2.42& 0.68& 0.59& 0.53& 0.54& 0.45\\
 & CodeT5 (9) & 5& 0.0& 0.23& 0.78& 0.66& 0.60& 0.61& 0.51\\
 & PLBART (3) & 9& 0.0& 7.48& 0.66& 0.59& 0.49& 0.49& 0.46\\
 & PLBART (6) & 5& 0.0& 0.10& 0.84& 0.73& 0.62& 0.66& 0.52\\
& CodeT5+220M (5) & 7& 0.0& 0.17& 0.74& 0.66& 0.62& 0.57& 0.47\\
& CodeT5+220Mbi (5) & 8& 0.0& 1.67& 0.64& 0.58& 0.51& 0.44& 0.44\\
& CodeT5+770M (12) & 5& 0.0& 0.05& 0.76& 0.69& 0.63& 0.59& 0.51\\
& CodeRL (12) & 5& 0.0& 0.13& 0.68& 0.62& 0.55& 0.56& 0.44\\
& CodeT5\_musu (5) & 7& 0.0& 2.17& 0.62& 0.55& 0.51& 0.48& 0.42\\
& CodeT5\_musu (12) & 7& 0.0& 0.50& 0.62& 0.61& 0.52& 0.48& 0.42\\
& CodeT5\_lntp (12) & 5& 0.0& 0.13& 0.66& 0.60& 0.55& 0.55& 0.43\\
& CodeT5\_lntp (24) & 5& 0.0& 0.13& 0.69& 0.64& 0.59& 0.55& 0.46\\
&Codegen (8) & 5& 0.0& 0.61& 0.70& 0.65& 0.54& 0.48& 0.59\\
&CodeT5+2B (10) & 5& 0.0& 0.21& 0.70& 0.70& 0.59& 0.51& 0.56\\
& CodeT5+2B\_dec (16) & 5& 0.0& 0.33& 0.55& 0.57& 0.48& 0.49& 0.48\\
& CodeT5+2B\_dec (32) & 5& 0.0& 0.54& 0.55& 0.57& 0.48& 0.49& 0.48\\
 \hline
\end{tabular}
\end{center}
\caption{Results of analysis by DirectProbe for tree distance prediction with 5 labels.}
\label{tab: dprobe_dist_ap}
\end{table*}

\begin{table*}[htb]
\tiny
\begin{center}
\begin{tabular}{llccccc}
\hline
\multicolumn{1}{c}{\bf Tokens}&\multicolumn{1}{c}{\bf Model (Layer)}  &\multicolumn{1}{c}{\bf No. of clusters}  &\multicolumn{2}{p{.78in}}{\bf Distance } &\multicolumn{2}{p{3cm}}{\bf Label Accuracy}\\\cline{4-5} \cline{6-7}
 & & &  Min & Avg & Not Siblings & Siblings\\
 \hline
 &CodeBERT (5) & 4& 0.19& 8.75& 0.87& 0.94\\
 & CodeBERT (9) & 4& 0.23& 8.55& 0.87& 0.93\\
 & CodeBERT (12) & 4& 0.18& 4.63& 0.87& 0.88\\
 &GraphCodeBERT (5) & 5& 0.24& 8.38& 0.87& 0.91\\
 & GraphCodeBERT (9) & 4& 0.24& 3.30& 0.84& 0.92\\
 \{Keyword-All\}& UniXcoder (5)  & 4& 0.20& 9.62& 0.86& 0.91\\
 & UniXcoder (9)  & 4& 0.14& 6.73& 0.80& 0.88\\
 & UniXcoder (12)  & 3& 0.0& 3.13& 0.61& 0.64\\
 &CodeT5 (5)  & 5& 0.17& 17.09& 0.84& 0.85\\
 & CodeT5 (9)  & 5& 0.70& 16.84& 0.86& 0.89\\
 & PLBART (3) & 4& 0.19& 14.17& 0.83& 0.86\\
  & PLBART (6) & 5& 0.58& 4.89& 0.88& 0.88\\
  & CodeT5+220M (5)  & 4& 0.04& 1.51& 0.91& 0.89\\
  & CodeT5+220Mbi (5)  & 5& 0.24& 4.56& 0.89& 0.82\\
  & CodeT5+770M (12)  & 4& 0.08& 1.55& 0.91& 0.91\\
  & CodeRL (12)  & 4& 0.21& 5.59& 0.89& 0.88\\
  & CodeT5\_musu (5)  & 5& 0.03& 5.56& 0.87& 0.83\\
  & CodeT5\_musu (12)  & 6& 0.0& 0.85& 0.80& 0.87\\
  & CodeT5\_lntp (12)  & 4& 0.19& 7.93& 0.89& 0.87\\
  & CodeT5\_lntp (24)  & 6& 0.0& 3.36& 0.83& 0.87\\
  & Codegen (8)  & 3& 1.76& 4.62& 0.79& 0.89\\
  & CodeT5+2B (10)  & 4& 0.64& 22.52& 0.84& 0.90\\
& CodeT5+2B\_dec (16)  & 3& 1.24& 3.83& 0.72& 0.86\\
  & CodeT5+2B\_dec (32)  & 5& 1.46& 15.88& 0.66& 0.74\\
  \hline
 &CodeBERT (5) & 7& 0.0& 6.68& 0.87& 0.91\\
 & CodeBERT (9) & 4& 0.31& 3.67& 0.88& 0.91\\
 & CodeBERT (12) & 3& 0.45& 8.55& 0.79& 0.87\\
 &GraphCodeBERT (5) & 4& 0.18& 0.81& 0.87& 0.92\\
 & GraphCodeBERT (9) & 4& 0.20& 4.33& 0.79& 0.91\\
 \{Keyword-Identifier\}& UniXcoder (5)  & 4& 0.13& 6.43& 0.82& 0.86\\
 & UniXcoder (9)  & 3& 0.11& 0.72& 0.76& 0.83\\
  & UniXcoder (12)  & 4& 0.14& 28.73& 0.47& 0.56\\
 &CodeT5 (5)  & 4& 0.16& 7.38& 0.76& 0.81\\
 & CodeT5 (9) & 4& 0.52& 19.72& 0.81& 0.85\\
 & PLBART (3) & 4& 0.13& 11.77& 0.78& 0.78\\
 & PLBART (6) & 4& 0.28& 5.17& 0.80& 0.87\\
  & CodeT5+220M (5)  & 3& 0.01& 1.63& 0.82& 0.82\\
  & CodeT5+220Mbi (5)  & 6& 0.0& 5.02& 0.61& 0.76\\
  & CodeT5+770M (12)  & 3& 0.05& 2.60& 0.83& 0.88\\
  & CodeRL (12)  & 3& 0.13& 5.55& 0.75& 0.80\\
  & CodeT5\_musu (5)  & 3& 0.0& 8.00& 0.69& 0.72\\
  & CodeT5\_musu (12)  & 3& 0.08& 2.94& 0.66& 0.75\\
  & CodeT5\_lntp (12)  & 3& 0.13& 5.06& 0.74& 0.78\\
  & CodeT5\_lntp (24)  & 4& 0.0& 0.68& 0.72& 0.79\\
  & Codegen (8)  & 2& 0.0& 0.0& 0.77& 0.85\\
  & CodeT5+2B (10)  & 3& 0.59& 3.68& 0.75& 0.84\\
& CodeT5+2B\_dec (16)  & 4& 1.44& 159.56& 0.78& 0.83\\
  & CodeT5+2B\_dec (32)  & 4& 2.56& 16.33& 0.67& 0.72\\
 \hline
\end{tabular}
\end{center}
\caption{Results of analysis by DirectProbe for siblings prediction with 2 labels.}
\label{tab: dprobe_sib_ap}
\end{table*}

\begin{table*}[htb]
\tiny
\begin{center}
\begin{tabular}{llcccccc}
\hline
\multicolumn{1}{c}{\bf Tokens}&\multicolumn{1}{c}{\bf Model (Layer)}  &\multicolumn{1}{p{1.2cm}}{\bf No. of clusters}  &\multicolumn{2}{c}{\bf Distance } &\multicolumn{2}{c}{\bf Label Accuracy}\\\cline{4-5} \cline{6-8}
 & & &  Min & Avg & No Edge & ComesFrom & ComputedFrom \\
 \hline
 &CodeBERT (5) & 5& 0.36& 7.59& 0.70& 0.95& 0.94\\
 & CodeBERT (9) & 5& 0.42& 7.54& 0.70& 0.95& 0.94\\
  & CodeBERT (12) & 4& 0.24& 3.68& 0.69& 0.91& 0.90\\
 &GraphCodeBERT (5) & 4& 0.41& 2.32& 0.68& 0.94& 0.94\\
 & GraphCodeBERT (9) & 4& 0.51& 2.90& 0.73& 0.95& 0.95\\
 \{Identifier-Identifier\}& UniXcoder (5) & 4& 0.41& 4.89& 0.66& 0.93& 0.91\\
 & UniXcoder (9) & 4& 0.34& 4.20& 0.64& 0.90& 0.88\\
 & UniXcoder (12) & 4& 0.92& 12.71& 0.54& 0.72& 0.79\\
 &CodeT5 (5) & 6& 0.0& 3.40& 0.69& 0.92& 0.81 \\
 & CodeT5 (9) & 4& 1.57& 15.00& 0.63& 0.90& 0.91\\
 & PLBART (3) & 6& 0.0& 4.76& 0.68& 0.90& 0.83\\
 & PLBART (6) & 4& 0.72& 8.99& 0.62& 0.91& 0.94\\
 & CodeT5+220M (5) & 4& 0.06& 1.47& 0.75& 0.89& 0.86\\
 & CodeT5+220Mbi (5) & 3& 0.18& 0.61& 0.70& 0.86& 0.79\\
 & CodeT5+770M (12) & 4& 0.11& 1.81& 0.74& 0.89& 0.89\\
 & CodeRL (12) & 5& 0.30& 7.19& 0.70& 0.85& 0.81\\
 & CodeT5\_musu (5) & 5& 0.0& 6.91& 0.71& 0.82& 0.79\\
 & CodeT5\_musu (12) & 4& 0.15& 2.29& 0.57& 0.81& 0.81\\
 & CodeT5\_lntp (12) & 4& 0.27& 4.98& 0.71& 0.85& 0.81\\
 & CodeT5\_lntp (24) & 4& 0.33& 3.65& 0.70& 0.87& 0.88\\
 & Codegen (8) & 4& 2.57& 26.09& 0.52& 0.82& 0.90\\
 & CodeT5+2B (10) & 4& 1.38& 21.53& 0.63& 0.88& 0.90\\
   & CodeT5+2B\_dec (16) & 4& 1.27& 13.04& 0.45& 0.78& 0.93\\
  & CodeT5+2B\_dec (32) & 5& 0.0& 7.76& 0.48& 0.80& 0.87\\
  \hline
\end{tabular}
\end{center}
\caption{Results of analysis by DirectProbe for data flow edge prediction with 3 labels.}
\label{tab: dprobe_dfg_ap}
\end{table*}

\begin{table*}[htb]
\tiny
\begin{center}
\begin{subtable}[t]{0.9\textwidth}
\hrule
    \begin{tabular}{p{1.8cm}lcccccccccc}
    & Cluster& 0& 1& 2& 3& 4& 5& 6& 7& 8& 9\\
    CodeBERT& Label& 3 & 2& 3& 5& 2& 3& 6& 6& 4& 5\\
    & Size& 178& 806& 453& 225& 241& 400& 683& 357& 1042& 815\\
\end{tabular}  
\hrule
\end{subtable}
\hfill
\begin{subtable}[t]{0.9\textwidth}
    \begin{tabular}{p{1.8cm}lccccccccc}
    & Cluster& 0& 1& 2& 3& 4& 5& 6& 7& 8\\
    GraphCodeBERT& Label&  2& 3& 5& 3& 2& 6& 5& 6& 4\\
    & Size& 48& 386& 94& 645& 999& 921& 946& 119& 1042\\
\end{tabular}  
\hrule
\end{subtable}
\hfill
\begin{subtable}[t]{0.9\textwidth}
    \begin{tabular}{p{1.8cm}lccccccccccccc}
    & Cluster& 0& 1& 2& 3& 4& 5& 6& 7& 8& 9& 10& 11& 12\\
    UniXCoder& Label& 3& 4& 6& 4& 6& 3& 2& 2& 5& 4& 3& 5& 6\\
    & Size& 334& 377& 225& 337& 83& 168& 662& 385& 646& 328& 529& 394& 732\\
\end{tabular}
\hrule
\end{subtable}
\hfill
\begin{subtable}[t]{0.9\textwidth}
    \begin{tabular}{p{1.8cm}lcccccccccc}
    & Cluster& 0& 1& 2& 3& 4& 5& 6& 7& 8& 9\\
    CodeT5& Label&  5& 2& 3& 2& 3& 6& 5& 4& 5& 6\\
    & Size& 26& 653& 354& 394& 677& 156& 61& 1042& 953& 884\\
\end{tabular}
\hrule
\end{subtable}
\hfill
\begin{subtable}[t]{0.9\textwidth}
    \begin{tabular}{p{1.8cm}lccccccccc}
    & Cluster& 0& 1& 2& 3& 4& 5& 6& 7& 8\\
    PLBART& Label& 2& 2& 3& 3& 6& 5& 6& 4& 5\\
    & Size& 105& 942& 614& 417& 227& 183& 813& 1042& 857\\
\end{tabular}  
\hrule
\end{subtable}
\hfill
\begin{subtable}[t]{0.9\textwidth}
    \begin{tabular}{p{1.8cm}lccccccccccccc}
    & Cluster& 0& 1& 2& 3& 4& 5& 6& 7& 8& 9& 10\\
    CodeT5+220M& Label& 3& 2& 3& 3& 4& 5& 4& 5& 4& 6& 6\\   
    & Size& 548& 1045& 329& 156& 51& 34& 759& 1015& 223& 965& 75\\
\end{tabular}
\hrule
\end{subtable}
\hfill
\begin{subtable}[t]{0.9\textwidth}
    \begin{tabular}{p{1.8cm}lccccccccccccc}
    & Cluster& 0& 1& 2& 3& 4& 5& 6& 7& 8& 9& 10\\
    Codegen& Label& 3& 3& 2& 6& 3& 2& 5& 5& 4& 4& 6\\   
    & Size& 272& 131& 219& 204& 629& 840& 166& 865& 41& 997& 836\\
\end{tabular}
\hrule
\end{subtable}
\end{center}
\caption{Cluster size and label for last layer of models for tree distance prediction task}
\label{tab: cl_dist}
\end{table*}

\begin{table*}[htb]
\tiny
\begin{center}
\begin{subtable}[t]{0.9\textwidth}
\hrule
    \begin{tabular}{p{1.8cm}lccc}
    & Cluster& 0& 1& 2\\
    CodeBERT& Label& Sibling& Sibling& Non-sibling\\
    & Size& 411& 779& 1210\\ 
\end{tabular}  
\hrule
\end{subtable}
\hfill
\begin{subtable}[t]{0.9\textwidth}
    \begin{tabular}{p{1.8cm}lcccc}
        & Cluster& 0& 1& 2& 3\\
    GraphCodeBERT& Label& Sibling& Non-sibling& Non-sibling& Sibling\\
    & Size& 1& 53& 1157& 1189\\
\end{tabular}  
\hrule
\end{subtable}
\hfill
\begin{subtable}[t]{0.9\textwidth}
    \begin{tabular}{p{1.8cm}lcccc}
    & Cluster& 0& 1& 2& 3\\
    UniXcoder& Label& Non-sibling& Sibling& Non-sibling& Sibling\\
    & Size& 2& 1153& 1208& 37\\
\end{tabular}
\hrule
\end{subtable}
\hfill
\begin{subtable}[t]{0.9\textwidth}
    \begin{tabular}{p{1.8cm}lccccccc}
    & Cluster& 0& 1& 2& 3& 4& 5& 6\\
    CodeT5& Label& Sibling& Non-sibling& Non-sibling& Sibling& Sibling& Sibling& Non-sibling\\
    & Size& 664& 458& 135& 157& 365& 4& 617\\
\end{tabular}
\hrule
\end{subtable}
\hfill
\begin{subtable}[t]{0.9\textwidth}
    \begin{tabular}{p{1.8cm}lccccc}
    & Cluster& 0& 1& 2& 3& 4\\
    PLBART& Label& Sibling& Sibling& Non-sibling& Sibling& Non-sibling\\
    & Size& 610& 126& 33& 454& 1177\\
\end{tabular}  
\hrule
\end{subtable}
\hfill
\begin{subtable}[t]{0.9\textwidth}
    \begin{tabular}{p{1.8cm}lccc}
    & Cluster& 0& 1& 2\\
    CodeT5+220M& Label& Non-Sibling& Non-Sibling& Sibling\\
    & Size& 608& 597& 1195\\ 
\end{tabular}  
\hrule
\end{subtable}
\hfill
\begin{subtable}[t]{0.9\textwidth}
    \begin{tabular}{p{1.8cm}lcccc}
    & Cluster& 0& 1& 2& 3\\
    Codegen& Label& Sibling& Non-Sibling& Sibling& Non-sibling \\
    & Size& 428& 2& 794& 1176
\end{tabular}  
\hrule
\end{subtable}
\end{center}
\caption{Cluster size and label for last layer of models for siblings prediction task}
\label{tab: cl_sib}
\end{table*}

\begin{table*}[!t]
\tiny
\begin{center}
\begin{subtable}[t]{.9\textwidth}
\hrule
    \begin{tabular}{p{1.8cm}lcccc}
    & Cluster& 0& 1& 2& 3\\
    CodeBERT& Label& NoEdge& NoEdge& Comes& Computed\\
    & Size& 1& 1208& 1206& 1185\\ 
\end{tabular}  
\hrule
\end{subtable}
\hfill
\begin{subtable}[t]{0.9\textwidth}
    \begin{tabular}{p{1.8cm}lccccccc}
        & Cluster& 0& 1& 2& 3& 4& 5& 6\\
    GraphCodeBERT& Label& Computed& NoEdge& Computed& NoEdge& Computed& NoEdge& Comes\\
    & Size& 1& 1& 1008& 549& 176& 659& 1206\\
\end{tabular}  
\hrule
\end{subtable}
\hfill
\begin{subtable}[t]{0.9\textwidth}
    \begin{tabular}{p{1.8cm}lcccc}
    & Cluster& 0& 1& 2& 3\\
    UniXcoder& Label& NoEdge& Computed& NoEdge& Comes\\
    & Size& 1& 1185& 1208& 1206\\
\end{tabular}
\hrule
\end{subtable}
\hfill
\begin{subtable}[t]{0.9\textwidth}
    \begin{tabular}{p{1.8cm}lcccc}
    & Cluster& 0& 1& 2& 3\\
    CodeT5& Label& NoEdge& Computed& NoEdge& Comes\\
    & Size& 1& 1185& 1208& 1206\\
\end{tabular}
\hrule
\end{subtable}
\hfill
\begin{subtable}[t]{0.9\textwidth}
    \begin{tabular}{p{1.8cm}lcccc}
    & Cluster& 0& 1& 2& 3\\
    PLBART& Label& NoEdge& Computed& NoEdge& Comes\\
    & Size& 1& 1185& 1208& 1206\\
\end{tabular}  
\hrule
\end{subtable}
\hfill
\begin{subtable}[t]{0.9\textwidth}
    \begin{tabular}{p{1.8cm}lccccc}
    & Cluster& 0& 1& 2& 3\\
    CodeT5+220M& Label& NoEdge& Computed& NoEdge& Comes \\
    & Size& 1& 1191& 1201& 1207\\
\end{tabular}  
\hrule
\end{subtable}
\hfill
\begin{subtable}[t]{0.9\textwidth}
    \begin{tabular}{p{1.8cm}lccccc}
    & Cluster& 0& 1& 2& 3& 4\\
    Codegen& Label& Computed& NoEdge& Computed& NoEdge& Comes \\
    & Size& 1145& 1126& 28& 101& 1200\\
\end{tabular}  
\hrule
\end{subtable}
\hfill
\end{center}
\caption{Cluster size and label for last layer of models for data flow edge prediction task}
\label{tab: cl_dfg}
\end{table*}

\end{document}